\documentclass[a4paper,fleqn,usenatbib]{mnras}

\usepackage{newtxtext,newtxmath}
\usepackage[T1]{fontenc}
\usepackage{ae,aecompl}

\usepackage{graphicx}	
\usepackage{amsmath}	
\usepackage{amssymb}	
\usepackage{lscape}
\usepackage{longtable}
\usepackage{xcolor}

\title[Stellar atmospheric parameters of FGK-type stars]{Stellar atmospheric parameters of FGK-type stars from high-resolution optical and near-infrared CARMENES spectra}

\author[E. Marfil et al.]{E.~Marfil,$^{1}$\thanks{E-mail: emigom01@ucm.es}
H.~M.~Tabernero,$^{2, 3}$
D.~Montes,$^{1}$
J.~A.~Caballero,$^{2}$
M.~G.~Soto,$^{4}$
\newauthor
J.~I.~Gonz\'{a}lez~Hern\'{a}ndez,$^{5, 6}$
A.~Kaminski,$^{7}$
E.~Nagel,$^{8}$
S.~V.~Jeffers,$^{9}$
A.~Reiners,$^{9}$
\newauthor
I.~Ribas,$^{10,11}$
A.~Quirrenbach,$^{7}$
P.~J. Amado$^{12}$
\\
$^{1}$Departamento de F{\'i}sica de la Tierra y Astrof{\'i}sica \& IPARCOS-UCM (Instituto de F\'{i}sica de Part\'{i}culas y del Cosmos de la UCM), \\Facultad de Ciencias F{\'i}sicas, Universidad Complutense de Madrid, 28040 Madrid, Spain\\
$^{2}$Centro de Astrobiolog\'{i}a (CSIC-INTA), ESAC, Camino Bajo del Castillo s/n, 28691, Villanueva de la Ca\~{n}ada, Madrid, Spain\\
$^{3}$Instituto de Astrof{\'i}sica e Ci{\^e}ncias do Espa\c{c}o, Universidade do Porto, CAUP, Rua das Estrelas, 4150-762 Porto, Portugal\\
$^{4}$School of Physics and Astronomy, Queen Mary, University of London, 327 Mile End Rd. London, United Kingdom\\
$^{5}$Universidad de La Laguna, Departamento de Astrof\'{i}sica, E-38206 La Laguna, Tenerife, Spain\\
$^{6}$Instituto de Astrof\'{i}sica de Canarias, v\'{i}a L\'{a}ctea s/n, 38205 La Laguna, Tenerife, Spain\\
$^{7}$Landessternwarte, Zentrum f\"{u}r Astronomie der Universit\"{a}t Heidelberg, K\"{o}nigstuhl 12, 69117 Heidelberg, Germany\\
$^{8}$Hamburger Sternwarte, Gojenbergsweg 112, 21029 Hamburg, Germany\\
$^{9}$Institut f\"{u}r Astrophysik, Georg-August-Universit\"{a}t-G\"{o}ttingen, Friedrich-Hund-Platz 1, D-37077 G\"{o}ttingen, Germany\\
$^{10}$Institut de Ci\`{e}ncies de l'Espai (CSIC), Campus UAB, C/ de Can Magrans s/n, 08193 Cerdanyola del Vall\`{e}s, Spain\\
$^{11}$Institut d'Estudis Espacials de Catalunya (IEEC), C/ Gran Capit\`{a} 2-4, 08034 Barcelona, Spain\\
$^{12}$Instituto de Astrof\'{i}sica de Andaluc\'{i}a (IAA-CSIC), Glorieta de la Astronom\'{i}a s/n, 18008 Granada, Spain
}

\date{Accepted XXX. Received YYY; in original form ZZZ}

\pubyear{2019}

\begin{document}
\label{firstpage}
\pagerange{\pageref{firstpage}--\pageref{lastpage}}
\maketitle

\begin{abstract}
With the purpose of assessing classic spectroscopic methods on high-resolution and high signal-to-noise ratio spectra in the near-infrared wavelength region, we selected a sample of 65 F-, G-, and K-type stars observed with CARMENES, the new, ultra-stable, double-channel spectrograph at the 3.5\,m Calar Alto telescope. We computed their stellar atmospheric parameters ($T_{\rm eff}$, $\log{g}$, $\xi$, and [Fe/H]) by means of the {\sc StePar} code, a Python implementation of the equivalent width method that employs the 2017 version of the MOOG code and a grid of MARCS model atmospheres. We compiled four \ion{Fe}{i} and \ion{Fe}{ii} line lists suited to metal-rich dwarfs, metal-poor dwarfs, metal-rich giants, and metal-poor giants that cover the wavelength range from 5\,300 to 17\,100\,{\small\AA}, thus substantially increasing the number of identified \ion{Fe}{i} and \ion{Fe}{ii} lines up to 653 and 23, respectively. We examined the impact of the near-infrared \ion{Fe}{i} and \ion{Fe}{ii} lines upon our parameter determinations after an exhaustive literature search, placing special emphasis on the 14 {\it Gaia} benchmark stars contained in our sample. Even though our parameter determinations remain in good agreement with the literature values, the increase in the number of \ion{Fe}{i} and \ion{Fe}{ii} lines when the near-infrared region is taken into account reveals a deeper $T_{\rm eff}$ scale that might stem from a higher sensitivity of the near-infrared lines to $T_{\rm eff}$.

\end{abstract}




\section{Introduction}
\label{sec:introduction}

The homogeneous, automated computation of stellar atmospheric parameters from stellar spectra, i.e. effective temperature $T_{\rm eff}$, surface gravity $\log{g}$, stellar metallicity [M/H], and micro-turbulent velocity $\xi$, plays a crucial role in many astrophysical contexts. First, it leads to the analysis of the fundamental properties of individual objects as well as of large stellar samples \citep{Val05, Adi14}. In this regard, large stellar spectroscopic surveys such as RAVE \citep{Ste06}, APOGEE \citep{All08}, the \emph{Gaia}-ESO Survey \citep{Gil12}, and GALAH \citep{DeS15} have laid the foundations for our current understanding of the structure and evolution of the Milky Way. Secondly, exoplanetary studies also rely on stellar parameter determinations not only to enable the determination of both planetary radii and masses \citep[e.g.][]{Man19, Sch19} but also to characterise the habitable zones around planet-harbouring stars \citep{Kas93, Kop13}. Furthermore, correlations between the stellar metallicity and planet occurrence rates are now well established and shed light on planet formation mechanisms \citep{Adi14, Mon18, Del18}.

The equivalent width ($EW$) method \cite[see e.g.][]{Sou08, Tab12, Tsa13, Muc13, Ben14, And16} is, along with the spectral synthesis method \cite[see e.g.][]{Val05, Pis17}, one of the most widely-used spectroscopic techniques for determining stellar atmospheric parameters. A full account of the key caveats of these two methods can be found in \citet{Jof19} and \citet{Bla19}. The advent of high-resolution near-infrared (NIR) spectrographs such as CARMENES \citep{Qui18}, SPIRou \citep{Art14}, GIANO \citep{Ori14, Oli18}, CRIRES+ \citep{Hat17}, IRD \citep{Kot14}, HPF \citep{Wri18}, and NIRPS \citep{Wil17} allows us to revisit these techniques, originally applied in the optical, in order to assess the impact of the NIR wavelength range on stellar parameter computations. In this context, new observations of FGK-type stars carried out with CARMENES\footnote{\tt http://carmenes.caha.es}, the double-channel spectrograph at the 3.5\,m Calar Alto telescope open up a unique opportunity to test the reliability of such techniques on high-resolution and high signal-to-noise (S/N) ratio spectra in the optical and near-infrared windows.

In this work, we compute the spectroscopic parameters of 65 FGK-type stars selected from a CARMENES stellar library by means of the $EW$ method, which relies on the strength (i.e the $EW$ measurements) of \ion{Fe}{i} and \ion{Fe}{ii} absorption lines to derive the stellar atmospheric parameters $T_{\rm eff}$, $\log{g}$, [Fe/H], and $\xi$ assuming local thermodynamic equilibrium (LTE). To do so, we followed the approach of \citet{Sou07} to automatically measure the $EW$ of the iron lines, and the {\sc StePar} code \citep{Tab19} to automatically compute the stellar atmospheric parameters imposing excitation and ionisation equilibrium conditions on the \ion{Fe}{i} and \ion{Fe}{ii} lines. 

The wavelength coverage provided by CARMENES, from 5\,200\,\AA\,up to 17\,100\,\AA, allowed us to substantially increase the number of \ion{Fe}{i} and \ion{Fe}{ii} lines subject to analysis with the $EW$ method with respect to previous studies restricted to the optical window \citep{Mel09, Jof14}. Furthermore, the high spectral resolution of CARMENES, which is $R$\,=\,94\,600 in the VIS channel and $R$\,=\,80\,400 in the NIR channel \citep{Qui18}, significantly improves both the line identification process and the $EW$ measurements. Despite the availability of iron line lists optimised for the NIR region in the literature, the impact on stellar parameter determinations of FGK-type stars is still unknown, mostly due to the fact that such line lists have not as yet been systematically applied to significantly large samples covering a wide portion of the stellar parameter space. For instance, \citet{And16} compiled a line list of \ion{Fe}{i} and \ion{Fe}{ii} lines in the region 10\,000--25\,000\,\AA, but only tested it against the spectra of the Sun and the F8~IV star HD~20010.

Several other spectral libraries of high-resolution spectra in the near-infrared have been developed over the past few years. For example, \citet{Leb12} presented the CRIRES-POP spectral library, which provides high-resolution ($R\sim$ 100\,000) spectra for 25 stars between B and M spectral types at 1--5 $\mu$m. Furthermore, \citet{Nic17} described the data reduction process and presented the first CRIRES-POP spectral atlas of the K giant 10~Leo. Although the resolution of the spectra in this library is comparable to that of CARMENES, the number of available spectra is significantly lower than the size of the library analysed in this work, and does not satisfactorily cover the parameter space of FGK-type stars. Another example is the IGRINS spectral library \citep{Par18}, which contains spectra of 84 stars between O and M spectral types in the $H$ (1.49--1.80\,$\mu$m) and $K$ (1.96--2.46\,$\mu$m) bands with a resolution of $R$\,=\,45\,000, which is almost half of that provided by CARMENES in the NIR channel. Finally, large surveys such as APOGEE \citep{Zam15, Maj17} have obtained intermediate-resolution ($R\sim$ 22\,500) spectra for hundreds of thousands of stars, but with a narrow wavelength coverage in the $H$ band (1.5--1.7\,$\mu$m).

The analysis performed in this work is structured as follows. In Sect.~\ref{sec:sample} we describe the selection of the sample. In Sect.~\ref{sec:analysis} we outline the main steps of our analysis, including the line selection process and the workflow of the {\sc StePar} code. In Sects.~\ref{sec:results} and \ref{sec:conclusions} we discuss the results and highlight the conclusions, respectively.

\section{Sample}
\label{sec:sample}

We observed an extensive sample of dwarf, giant, and supergiant stars and brown dwarfs with spectral types from O4 to late L as part of the first open time proposal that used CARMENES. While further details on this stellar library will be provided in forthcoming publications, we start here its scientific exploitation.

From the stellar library we selected 65 stars with spectral types later than F5 and earlier than K4, and projected equatorial rotational velocities $\varv\sin{i}<15$\,km\,s$^{-1}$ (see Table~\ref{tab:par_stars_ref}). The restriction in spectral type stems from the general limitations of the $EW$ method and hence, {\sc StePar}, as explained in \citet{Tab19}, while stars with high rotational velocities have line profiles that cannot be properly fitted by a Gaussian shape, leading to less reliable $EW$ measurements. None of the observed 65 FGK-type stars had a known visual (physical) or optical (non-physical) companion at less than 5\,arcsec. However, we excluded from this analysis one of the giants found in the library, c~Gem, with spectral type K4.5~III \citep{Kee89}, as it appeared as an SB2 binary system after cross-correlating its spectrum with the atlas spectrum of Arcturus, as explained in Sect.~\ref{subsec:pro}. 

Our target list contains 14 \emph{Gaia} benchmark stars \citep{Jof14, Jof18, Hei15}, including the Sun. The spectrum of the Sun was obtained through the observation of the asteroid 1~Ceres thanks to the allocation of Calar Alto Director's discretionary time. According to their original purpose, the fact that the fundamental parameters of these stars have been computed independently from spectroscopy makes them suitable as a reference to assess any method aimed at the automated analysis of cool stars.

Table~\ref{tab:par_stars_ref} displays the star names, Henry-Draper numbers, equatorial coordinates from 2MASS \citep{Skr06}, parallaxes from the {\it Gaia} Data Release 2 \citep{gaia} if available, and the Hipparcos mission \citep{Lee07}, along with the spectral types, the values of $T_{\rm eff}$, $\log{g}$, $\xi$, [Fe/H] and the stellar projected rotational velocities, $\varv\sin{i}$, found in the literature for the selected sample. For the \emph{Gaia} benchmark stars, we adopted the parameters from \citet{Jof14} and \citet{Hei15}, with updated values from \citet{Jof18}. For the remaining stars, we tabulate the stellar parameters from the most recent references found in the PASTEL catalogue \citep{Sou16}.
   
\begin{figure}
\centering
\includegraphics{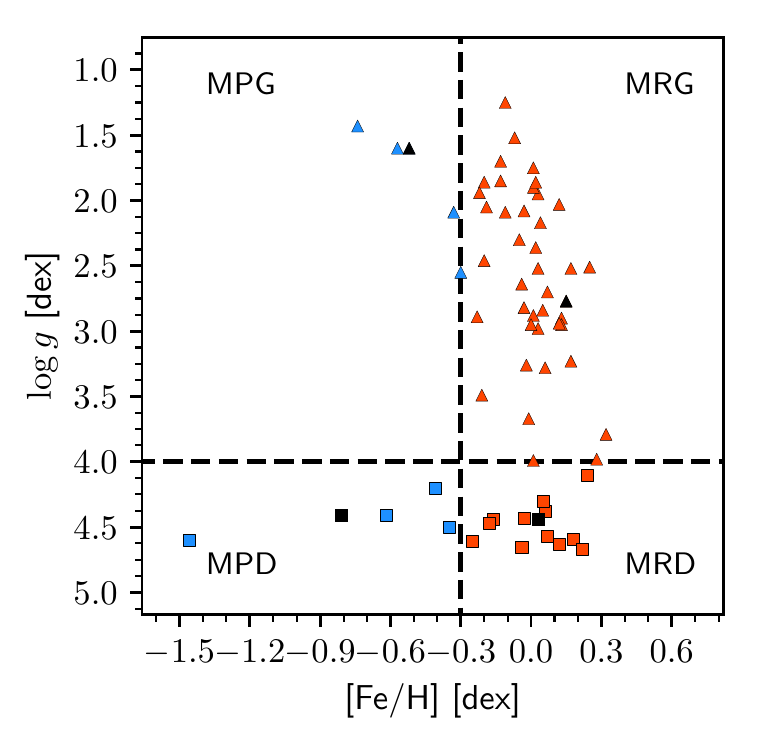}
\caption{Division of the parameter space in the sample according to the stellar atmospheric parameters found in the literature. Vertical and horizontal dashed black lines represent the boundaries at [Fe/H]$\,=-0.3$\,dex and $\log{g}$\,$=4.0$\,dex, respectively, for metal-rich dwarfs (MRD, orange squares), metal-poor dwarfs (MPD, blue squares), metal-rich giants (MRG, orange triangles), and metal-poor giants (MPG, blue triangles). The stars taken as a reference for each of these regions are shown in black.}
\label{fig:sample_histogram} 
\end{figure}

\begin{figure}
\centering
\includegraphics{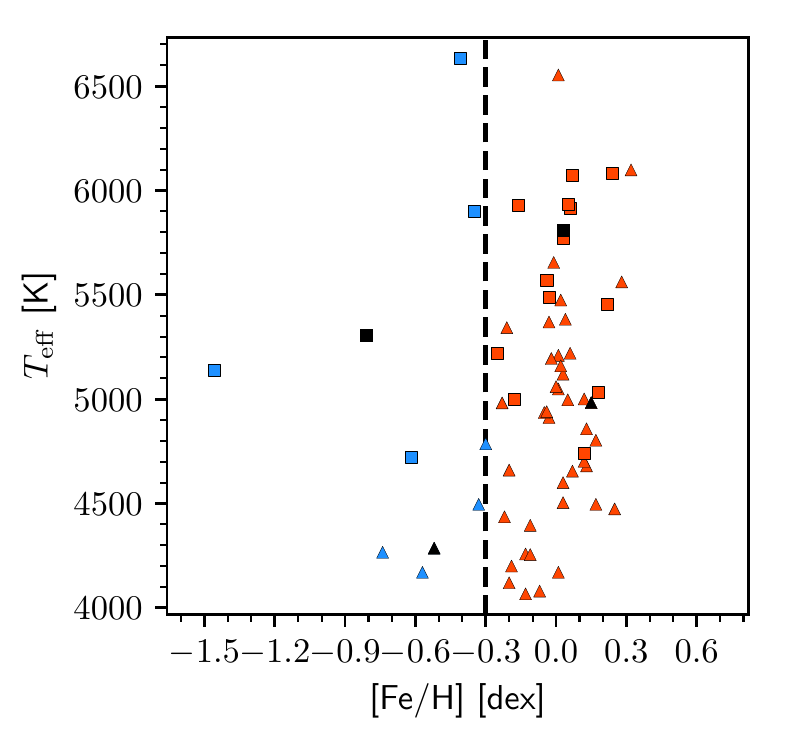}
\caption{Same as Fig.~\ref{fig:sample_histogram}, but for literature values of $T_{\rm eff}$ versus [Fe/H] in the sample. Only the boundary at [Fe/H]\,$=-0.3$\,dex is shown.}
\label{fig:sample_histogram_teff} 
\end{figure}

Following \citet{Tab19}, we divided the parameter space into four different regions in terms of $\log{g}$ and [M/H], using [Fe/H] as a proxy of stellar metallicity, in order to simplify our search for iron lines in the CARMENES spectra, as explained in Sect.~\ref{sect:line_selections}. We thus made a distinction between the dwarf regime, $\log{g} \ge 4.00$, and the giant regime, $\log{g} < 4.00$, and between metal-rich stars, [Fe/H] > $-$0.30, and metal-poor stars, [Fe/H] $\le$ $-$0.30. We dubbed the four resulting line lists metal-rich dwarfs (MRD), metal-poor dwarfs (MPD), metal-rich giants (MRG) and metal-poor giants (MPG). We selected the following \emph{Gaia} benchmark stars, all of which were observed with CARMENES, as a reference for the assembly of the corresponding \ion{Fe}{i} and \ion{Fe}{ii} line lists: 18 Sco for the MRD, $\mu$ Cas for the MPD, $\epsilon$ Vir for the MRG, and Arcturus for the MPG. We show this division of the parameter space in Figs.~\ref{fig:sample_histogram} and \ref{fig:sample_histogram_teff}.

\section{Analysis}
\label{sec:analysis}

\subsection{Data processing}
\label{subsec:pro}

The 65 pairs of VIS and NIR spectra were taken in service mode between March and June 2016 with the two CARMENES channels operating simultaneously. In general, exposure times were manually adjusted to reach a signal-to-noise ratio (S/N) between 100 and 300 in the $J$ band. The observations were carried out without the simultaneous wavelength calibration of the Fabry-P\'{e}rot etalons since there was no particular interest in precise radial velocity determinations (i.e. better than $\sim$ 20~m~s$^{-1}$) for these stars. 

The spectra were taken in ``target+sky'' mode, that is, the stars were observed in fibre A and the sky in fibre B. Both fibres are identical but fibre B is located at 88\,arcsec to the east. Star and sky spectra are available through the Calar Alto archive. In our work we did not subtract the corresponding sky spectrum to each star spectrum, as this is an on-going analysis \citep{Nag20}.

\begin{figure}
\centering
\includegraphics{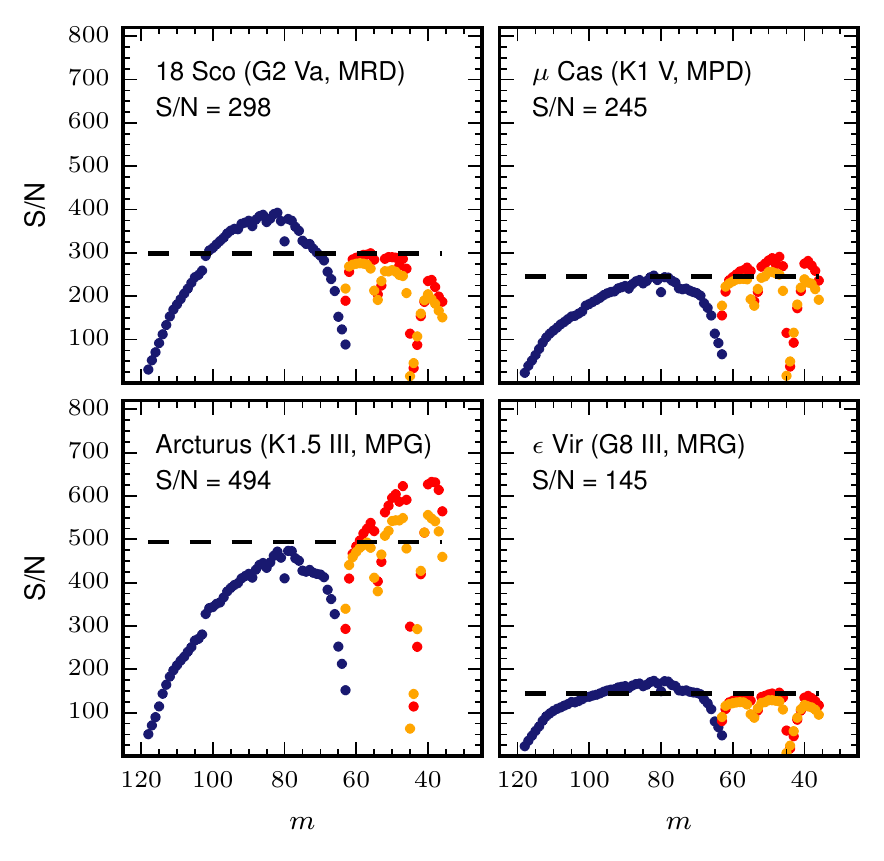}
\caption{\label{fig:SNR_orderwise} CARACAL S/N of the CARMENES spectra of the reference stars (18 Sco, $\mu$~Cas, $\epsilon$~Vir, and Arcturus) as a function of the spectral order $m$. The blue circles are the orders in the VIS channel, while the orange and red circles are the two HgCdTe array detectors of the NIR channel. The dashed black lines mark the global S/N estimation given by iSpec.}
\end{figure}

The raw spectra were reduced with the CARACAL pipeline \citep{Zec14, Cab16}, which is based on the IDL REDUCE package \citep{Pis02}. CARACAL generates one fully reduced, wavelength-calibrated, one-dimensional spectrum of the individual spectral orders.  Fig.~\ref{fig:SNR_orderwise} displays the CARACAL S/N of the four reference spectra as a function of the diffraction order $m$. We estimated the global S/N of the spectra with the integrated Spectroscopic framework \citep[iSpec, see][]{Bla14} in terms of the median of the flux values divided by their corresponding flux errors. The global S/N of the selected spectra can also be found in Table~\ref{tab:par_stars_stepar}.

Next, we employed a wavelength grid to merge the spectral orders of both channels into one single spectrum. The wavelength grid, which is evenly spaced on a logarithmic scale, mirrors the natural wavelength spacing of the CARMENES spectrographs across the orders. In Fig.~\ref{fig:spectral_lines} we show the normalised, merged spectra of the four stars taken as a reference in this work.

Since the CARMENES instrument operates in vacuum, we performed a vacuum-to-air wavelength conversion of the order-merged, channel-merged, CARMENES spectra to provide the wavelengths of the \ion{Fe}{i} and \ion{Fe}{ii} lines on an air scale, following the International Astronomical Union standard \citep{Mor00}:
\begin{equation}
\lambda_{\rm air} = \frac{\lambda_{\rm vacuum}}{n},
\end{equation}
where $n$ is the refraction index, which is given by the following expression:
\begin{equation}
n = 1 + 8.34254\times10^{-5} + \frac{2.406147\times10^{-2}}{130 - s^2} + \frac{1.5998\times10^{-4}}{38.9 - s^2},
\end{equation}
where $s=10^4/\lambda_{\rm vacuum}$, with $\lambda_{\rm vacuum}$ in {\AA}.

\begin{figure}
\centering
\includegraphics{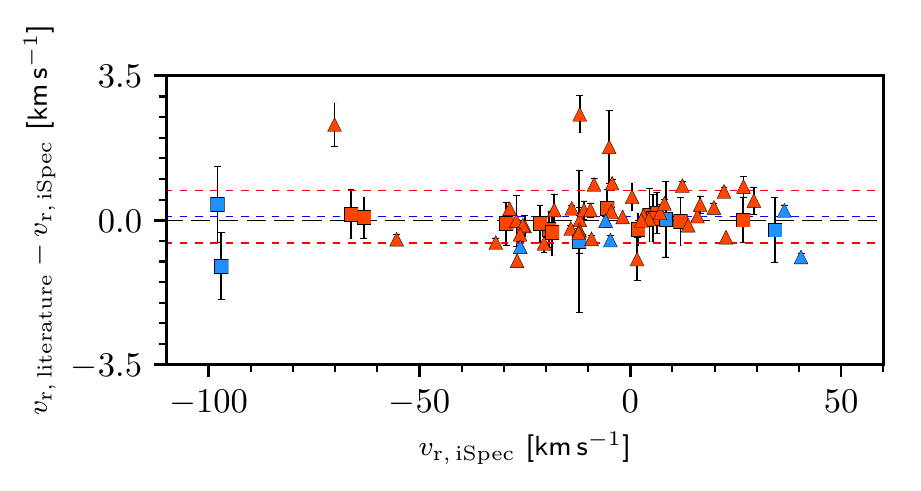}
\caption{\label{fig:rvsample} Comparison between the radial velocities $\varv_{\rm r}$ of the sample obtained with iSpec and the literature values. Symbols are the same as in Fig.~\ref{fig:sample_histogram}. The dotted blue and red lines are the average difference and the corresponding 1$\sigma$ dispersion, respectively.}
\end{figure}

After the vacuum-to-air wavelength conversion, we accounted for the barycentric velocity of the observatory at the time of observations. We then computed the radial velocities with iSpec by means of the cross-correlation function between the observed CARMENES spectra and a template spectrum provided by iSpec in the following way. In the dwarf regime, we set as the template a solar spectrum based on data from the NARVAL \citep{Aur03} and HARPS \citep{May03} instruments \citep[see][]{Bla14} covering the overlap region with CARMENES, i.e. the 5\,200--10\,480\,{\AA} range. Likewise, in the giant regime we set as the template spectrum an atlas of Arcturus covering the 5\,200--9\,260\,{\AA} range \citep{Hin00}. Both template spectra were corrected from telluric absorption features, which makes them suitable for cross correlation. This allowed us to correct the spectra from the corresponding Doppler shift. In Fig.~\ref{fig:rvsample} we compare the radial velocities thus computed against the literature values. Four stars exhibit a difference in radial velocity greater than 1~km\,s$^{-1}$ compared to literature values. These are all singled-lined (SB1) spectroscopic binaries: $\mu$~Cas \citep{Wor77}, $\alpha$~CMi \citep{Gira00}, $\alpha$~UMa \citep{Spe37}, and $\zeta$~Her \citep{Sca83}. The radial velocities of our sample can also be found in Table~\ref{tab:par_stars_stepar}. The average difference in the computed radial velocities of the sample with respect to the literature values is 0.09\,$\pm$\,0.64~km\,s$^{-1}$.

\subsection{Fe\,{\sc i} and Fe\,{\sc ii} line selections} \label{sect:line_selections}

\begin{figure*}
\includegraphics{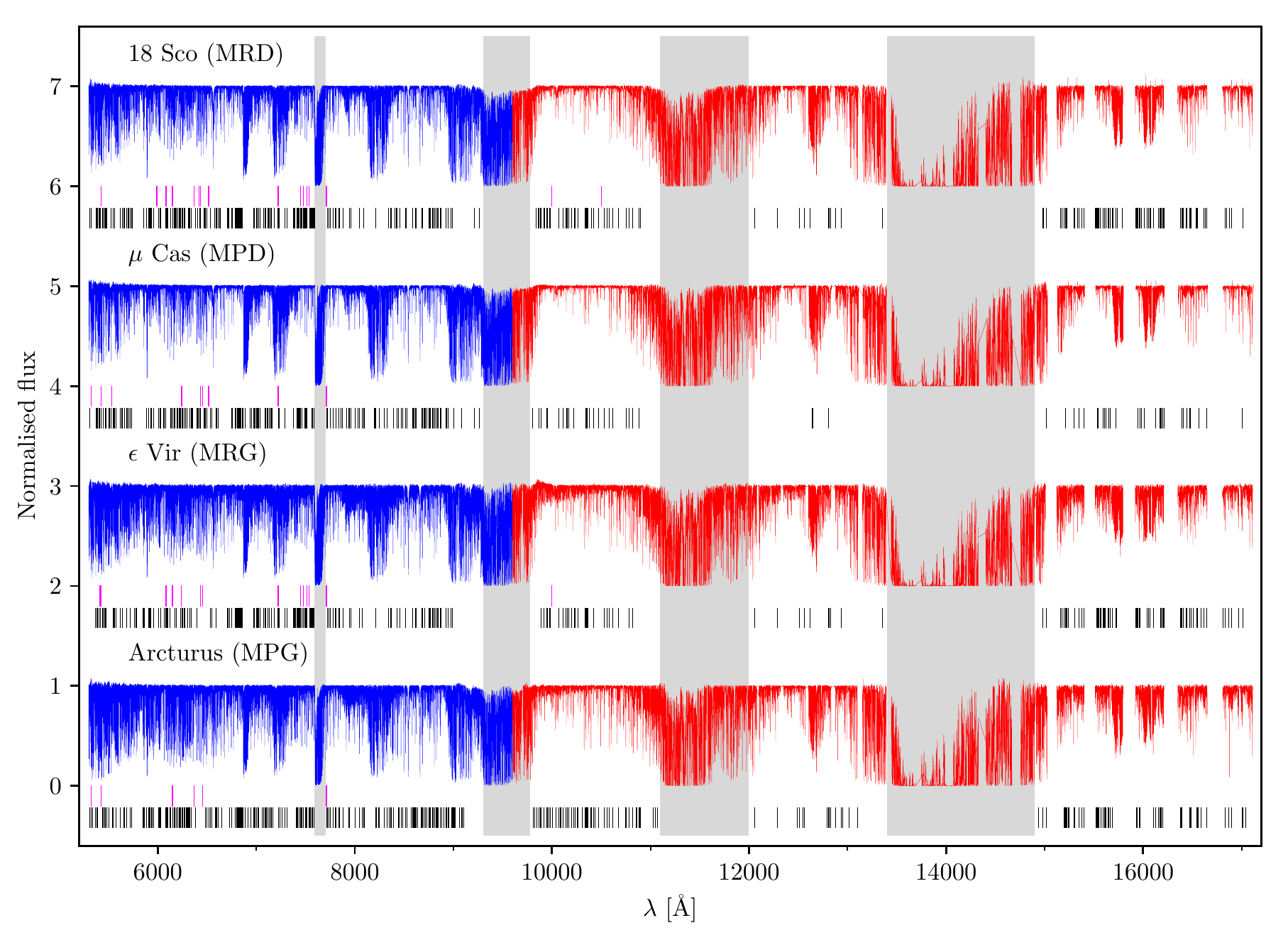}
\caption{\label{fig:spectral_lines} Distribution of the selected \ion{Fe}{i} and \ion{Fe}{ii} absorption lines in the reference spectra. The \ion{Fe}{i} and \ion{Fe}{ii} lines are shown as black and pink vertical lines, respectively, below the spectra. The VIS and NIR channels of the CARMENES instrument are shown in blue and red, respectively. The grey shaded areas show the regions severely affected by telluric absorption.}
\end{figure*}

\begin{table}
\centering
\caption{\label{tab:nlines} Number of \ion{Fe}{i} and \ion{Fe}{ii} lines reported in this work, \citet[][Sou08]{Sou08}, \citet[][And16]{And16}, and \citet[][Tab19]{Tab19} from 5300 to 17100\,{\AA}.}
\begin{tabular}{llcc}
\hline\noalign{\smallskip}
Reference & Line list/region & \multicolumn{2}{c}{\#lines}\\
\hline\noalign{\smallskip}
& & \ion{Fe}{i} & \ion{Fe}{ii}\\
\hline\noalign{\smallskip}
This work & MRD                  & 386 & 16 \\
This work & MPD                  & 295 &  9 \\
This work & MRG                  & 306 & 13 \\
This work & MPG                  & 379 &  4 \\
This work & CARMENES VIS channel & 437 & 21 \\
This work & CARMENES NIR channel & 216 &  2 \\
This work & Globally             & 653 & 23 \\
Tab19 & MRD                      & 112 &  8 \\
Tab19 & MPD                      &  82 &  8 \\
Tab19 & MRG                      &  72 &  7 \\
Tab19 & MPG                      &  95 &  5 \\
Tab19 & Globally                 & 175 & 14 \\
Sou08 & ...                      & 172 & 19 \\
And16 & ...                      & 272 & 12 \\
\hline
\end{tabular}
\end{table}

We requested four line lists from the Vienna Atomic Line Database (VALD3; \citealt{Rya15, Kup00, Kup99, Pis95}), corresponding each to one of our four reference spectra. We used the option {\tt Extract stellar} available at the VALD3 website\footnote{\tt{http://vald.astro.uu.se}}, with a wavelength range from 5\,300 to 17\,100\,{\AA}, a minimum line depth of 5\% with respect to the continuum flux, and the corresponding input stellar parameters found in Table~\ref{tab:par_stars_ref}. We excluded the wavelength range 5\,200--5\,300\,{\AA} from this search because of the low S/N of the CARMENES spectra in this region.

Because of its user-friendly interface, we used iSpec to select the \ion{Fe}{i} and \ion{Fe}{ii} spectral lines by visually projecting the VALD3 line list files onto the corresponding processed reference spectra. We rejected Fe~{\sc i} and Fe~{\sc ii} lines that showed spectral blending with close atomic and molecular lines. Since telluric lines are ubiquitous in the near-infrared and at the red end of the optical \citep[see e.g.][]{Rei18}, we computed a synthetic transmission spectrum via the telluric-correction tool {\tt molecfit} \citep{Kau15, Sme15}, which makes use of the line-by-line radiative transfer model \citep[LBLRTM,][]{Clo05} and the HITRAN molecular line database \citep{Gor17}, to model the Earth's atmospheric transmission spectrum. This allowed us to prevent wrong line identification throughout the visual inspection of the reference spectra. Further details on the telluric correction of the CARMENES spectra can be found in \citet{Pas19}. A full description of the correction will appear in a forthcoming publication of the CARMENES series \citep{Nag20}.

To expedite our analysis, we also looked for \ion{Fe}{i} and \ion{Fe}{ii} line compilations found in the literature that overlap with the wavelength range covered by CARMENES. Since the careful analysis of the optical wavelength range up to $\sim$6\,860\,{\AA} has already led to several line lists published in previous works that were specifically compiled to yield the best possible set of stellar atmospheric parameters for FGK-type stars \citep[see e.g.][]{Sou08, Jof14, Tab19}, we refrained from further refining the line selection in this window and adopted the iron lines given in \citet{Sou08}. As to the near-infrared region, we checked our iron line selections from 10\,000 to 17\,100\,{\AA} against the ones tabulated in \citet{And16}. Despite our careful search for \ion{Fe}{ii} in the NIR region, we only found one \ion{Fe}{ii} line at $\lambda$\,=\,10\,501.503\,{\AA}. Finally, iron lines found in the region 6\,800--10\,000\,{\AA} were not compared with the literature due to the lack of line compilations in this spectral window. In Table~\ref{tab:nlines} we show a summary of the number of iron lines listed in this work on a global and per-line list basis, i.e. MRD, MPD, MRG, and MPG, in comparison with those tabulated in \citet{Sou08} and \citet{And16} in the wavelength region covered by CARMENES.

Since we assembled the line lists considering four specific reference spectra, we removed the \ion{Fe}{i} and \ion{Fe}{ii} line identifications that fall into any of the CARMENES inter- and intra-order gaps\footnote{\tt http://carmenes.caha.es/ext/instrument/} as a consequence of the corresponding Doppler shift corrections in the remaining spectra of the sample.

In Fig.~\ref{fig:spectral_lines} we show the distribution of the selected \ion{Fe}{i} and \ion{Fe}{ii} lines in the reference spectra. In addition, in Fig.~\ref{fig:18Scospectrum} we give a close-up view of the spectrum of the reference, solar-type star 18~Sco along with the line selections. We give the central wavelength in air, $\lambda_{\rm air}$, the excitation potential, $\chi$, and the oscillator strength, $\log{gf}$, of the selected \ion{Fe}{i} and \ion{Fe}{ii} lines in Tables~\ref{tab:line_table_all_fe_i} and \ref{tab:line_table_all_fe_ii}, respectively.

\subsection{EW measurements} \label{ewmeasurements}

\begin{figure*}
\centering
\includegraphics[width=\textwidth]{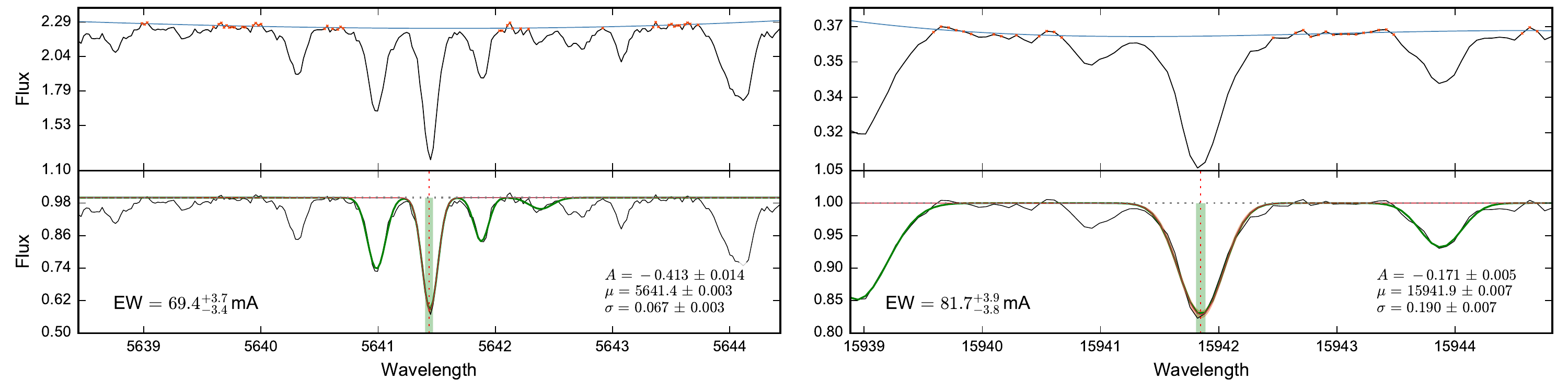}
\caption{\label{fig:ew} Equivalent width measurements of two \ion{Fe}{i} lines in the spectrum of 18~Sco, at 5\,641.434 {\AA} (left) and 12\,824.859 {\AA} (right). The upper panels illustrate the continuum determination, where the points used for the final polynomial fit are highlighted in red. The bottom panels show the full fit performed for all detected lines, shown in green, and the Gaussian fit of the selected line, shown in red, parametrised by the central intensity in normalised units, $A$, the central wavelength in {\AA}, $\mu$, and the Gaussian dispersion, $\sigma$. The shaded red area depicts the 1$\sigma$ confidence intervals of the Gaussian fit, and the green square, the $EW$ estimation, as explained in the text.}
\end{figure*}

We computed the $EW$s by fitting Gaussian profiles to the absorption lines\footnote{The code is available at: \\{\tt https://github.com/msotov/EWComputation}.}, as shown in Fig~\ref{fig:ew}. First, we selected a region approximately 6\,{\AA} wide centred at the selected absorption line, $l$, and performed a continuum  normalisation on the spectra following \citet{Sou07}. Specifically, we fitted a third degree polynomial to the data, selecting only the points that lie within {\tt rejt} times the polynomial, where {\tt rejt}\,$ = 1-1/{\rm (S/N)}$, and S/N is the signal-to-noise ratio of the region. We then identified the absorption lines present in the spectra by finding the points where the first derivative of the data was zero, and the second derivative was positive. Finally, we fitted Gaussian profiles to the lines detected, and integrated the profile corresponding to the selected line $l$ to obtain the $EW$. The uncertainty in the $EW$ was estimated by changing the Gaussian parameter estimates within 1$\sigma$ of their uncertainty for a total of 1\,000 iterations, and looking at the $EW$ distribution.

As in \citet{Tab19}, we only considered lines with 10 m{\AA} < $EW$ < 120 m{\AA} for all stars in the sample to avoid problems with line profiles of very intense lines and potentially bad $EW$ measurements of extremely weak lines.

\subsection{\sc StePar} \label{sect:stepar}

The {\sc StePar} code\footnote{{\sc StePar} is available at: \\{\tt https://github.com/hmtabernero/StePar}.} is a Python implementation of the $EW$ method specifically designed for the automated and simultaneous computation of the stellar atmospheric parameters of FGK-type stars, namely $T_{\rm eff}$, $\log{g}$, [Fe/H], and $\xi$. {\sc StePar} is one of the thirteen pipelines in the \emph{Gaia}-ESO Survey used in the analysis of UVES U580 spectra of late-type, low-mass stars. A full description of its workflow and performance can be found in \citet{Tab19}. {\sc StePar} is an iterative code that derives the stellar parameters and their associated uncertainties by imposing both excitation and ionisation equilibrium conditions on a set of \ion{Fe}{i} and \ion{Fe}{ii} lines, using the 2017 version of the MOOG\footnote{\tt https://www.as.utexas.edu/\textasciitilde chris/moog.html} code \citep{Sne73} and a grid of plane-parallel and spherical MARCS\footnote{\tt http://marcs.astro.uu.se} model atmospheres \citep{Gus08}.

For any given MOOG-compliant $EW$ input file comprised of a significant number of \ion{Fe}{i} and \ion{Fe}{ii} lines, {\sc StePar} follows a Downhill Simplex minimisation algorithm \citep{Pre02} across the parameter space in order to find the stellar atmospheric parameters that best reproduce the observed $EW$s. The code takes $T_{\rm eff}$ = 5777~K, $\log{g}$ = 4.44~dex, and $\xi$ = 1.0~km\,s$^{-1}$ as the initial input values.

If we let $\epsilon({\rm Fe})$ represent the iron abundance retrieved from any given Fe line and $\chi$ be the excitation potential of the line, {\sc StePar} iterates until the slopes of $\chi$ vs. $\log{\epsilon({\rm Fe\,\textsc{i}})}$ and $\log{EW/\lambda}$ vs. $\log{\epsilon({\rm Fe\,\textsc{i}})}$ are zero, i.e. the iron atoms are in excitation equilibrium. It also imposes ionisation equilibrium so that $\log{\epsilon({\rm Fe\,\textsc{i}})}$ = $\log{\epsilon({\rm Fe\,\textsc{ii}})}$. Throughout this iterative process, the code verifies that the average [Fe/H] in the MOOG output is always compatible with the iron abundance of the input atmospheric model. Next, {\sc StePar} performs an individual $\sigma$ clipping on the \ion{Fe}{i} and \ion{Fe}{ii} lines to remove the ones that imply an iron abundance, $\log{\epsilon(\rm Fe)}$, that exceeds the 3$\sigma$ limit with respect to the median abundance of all lines. After this step, {\sc StePar} restarts the minimisation algorithm with the remaining \ion{Fe}{i} and \ion{Fe}{ii} lines, taking as initial input values the parameters computed in the first run. {\sc StePar} computes the uncertainties in the stellar atmospheric parameters following the sequence: $\delta\xi$, $\delta T_{\rm eff}$, $\delta\log{g}$, and $\delta$[Fe/H]. This computation relies on the retrieved Fe~{\sc i} and Fe~{\sc ii} abundances and the uncertainties in the slopes that define the equilibria conditions. The code also propagates the uncertainties following the previous sequence. For example, the uncertainty in [Fe/H] is a quadrature between the standard deviation of the Fe~{\sc i} and Fe~{\sc ii} abundances and the propagated uncertainties in the remaining stellar parameters. Further details on the computation of the uncertainties can be found in \citet{Tab19}.

\section{Results and discussion}
\label{sec:results}

\begin{figure}
\centering
\includegraphics{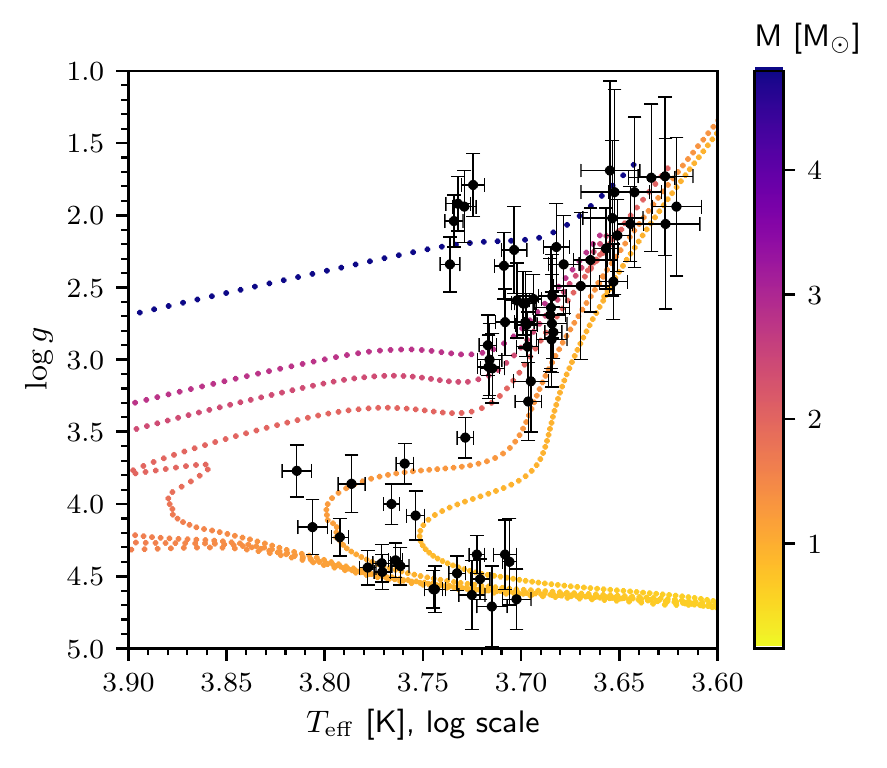}
\caption{Kiel diagram ($\log{g}$ vs. $\log{T_{\rm eff}}$) of the sample along with the YaPSI isochrones at 0.1, 0.4, 0.6, 1, 4, and 13 Ga (for Z=0.016, see \citealt{Spa17}).}
\label{fig:kiel_diagram}
\end{figure}

\begin{figure}
\centering
\includegraphics{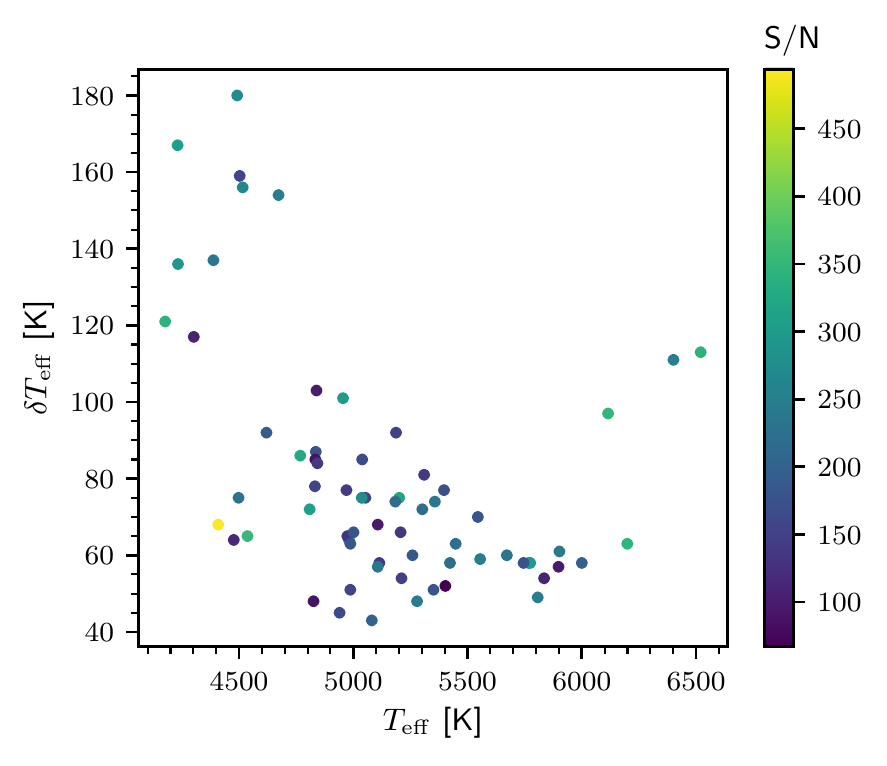}
\caption{Uncertainties in $T_{\rm eff}$, $\delta T_{\rm eff}$, versus $T_{\rm eff}$ for our sample, as computed with {\sc StePar}.}
\label{fig:T_eff_errors_all} 
\end{figure}

\begin{figure}
\centering
\includegraphics{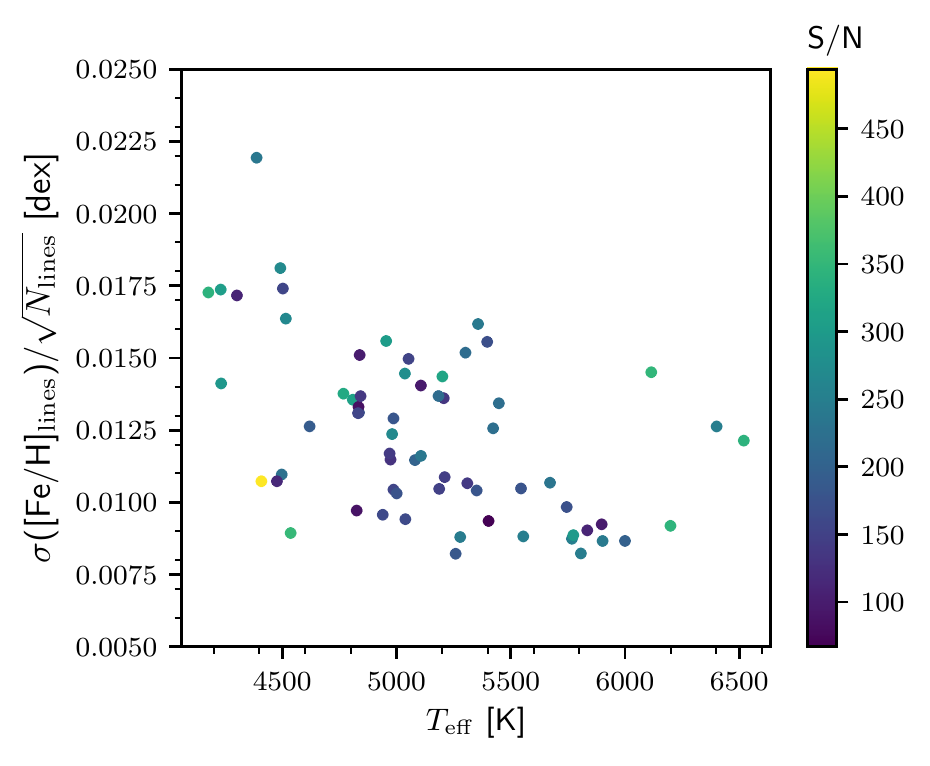}
\caption{\label{fig:fe_dispersion} Line-to-line scatter in [Fe/H] versus $T_{\rm eff}$ and S/N in the sample.}
\end{figure}

\begin{table}
\centering
\caption{Summary of the Monte Carlo simulations carried out on the $T_{\rm eff}$, $\log{g}$, and [Fe/H] values of the sample as computed with {\sc StePar}. We show the average difference on each parameter and the values of the Pearson ($r_{\rm p}$) and Spearman ($r_{\rm s}$) correlation coefficients.}
\label{tab:mc}
\begin{tabular}{cccc}
\hline
Parameter & Difference & $r_{\rm p}$ & $r_{\rm s}$\\
\hline
\multicolumn{4}{c}{VIS and NIR channels}\\
\hline
$T_{\rm eff}$ {[}K{]} & $-$100  $\pm$ 166  &    0.40 $\pm$ 0.07 &    0.41 $\pm$ 0.07 \\
$\log{g}$ {[}dex{]}   & $-$0.03 $\pm$ 0.38 &    0.10 $\pm$ 0.10 &    0.07 $\pm$ 0.11 \\
{[}Fe/H{]} {[}dex{]}  &    0.00 $\pm$ 0.11 & $-$0.09 $\pm$ 0.06 & $-$0.12 $\pm$ 0.07 \\
\hline
\multicolumn{4}{c}{VIS channel only}\\
\hline
$T_{\rm eff}$ {[}K{]}& $-$92 $\pm$ 135      &    0.21 $\pm$ 0.08 &    0.21 $\pm$ 0.09 \\
$\log{g}$ {[}dex{]}    & $-$0.01 $\pm$ 0.38 & $-$0.01 $\pm$ 0.10 &    0.00 $\pm$ 0.10 \\
{[}Fe/H{]} {[}dex{]}   & $-$0.04 $\pm$ 0.10 & $-$0.01 $\pm$ 0.08 & $-$0.07 $\pm$ 0.09 \\
\hline
\end{tabular}
\end{table}

\begin{figure*}
\centering
\includegraphics{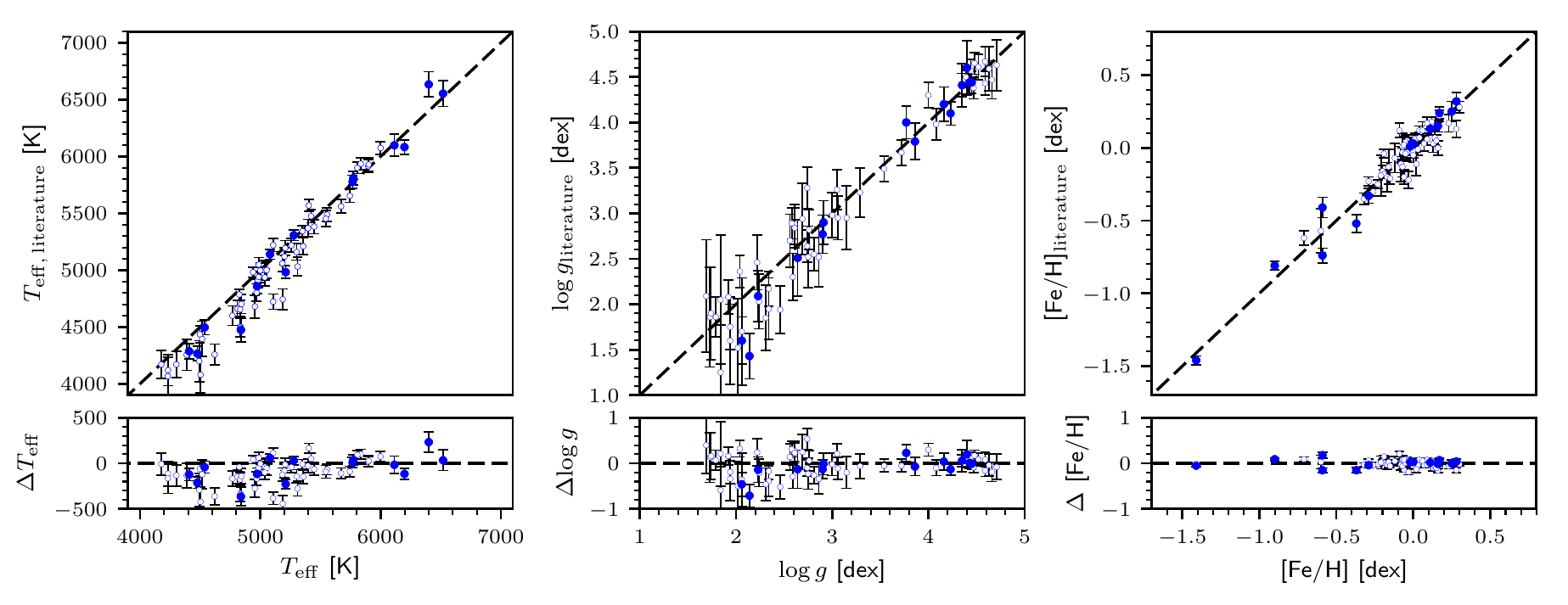}
\caption{\label{fig:Teff_logg_FeH_all} Comparison between the stellar atmospheric parameters obtained with {\sc StePar} including the VIS and NIR channels of CARMENES and the literature values. Blue filled circles are the \emph{Gaia} benchmark stars in our sample. The remaining stars in the sample are shown with blue open circles. Dashed black lines indicate the one-to-one relationship. From left to right: $T_{\rm eff}$, $\log{g}$, and [Fe/H].}
\end{figure*}

\begin{figure*}
\centering
\includegraphics{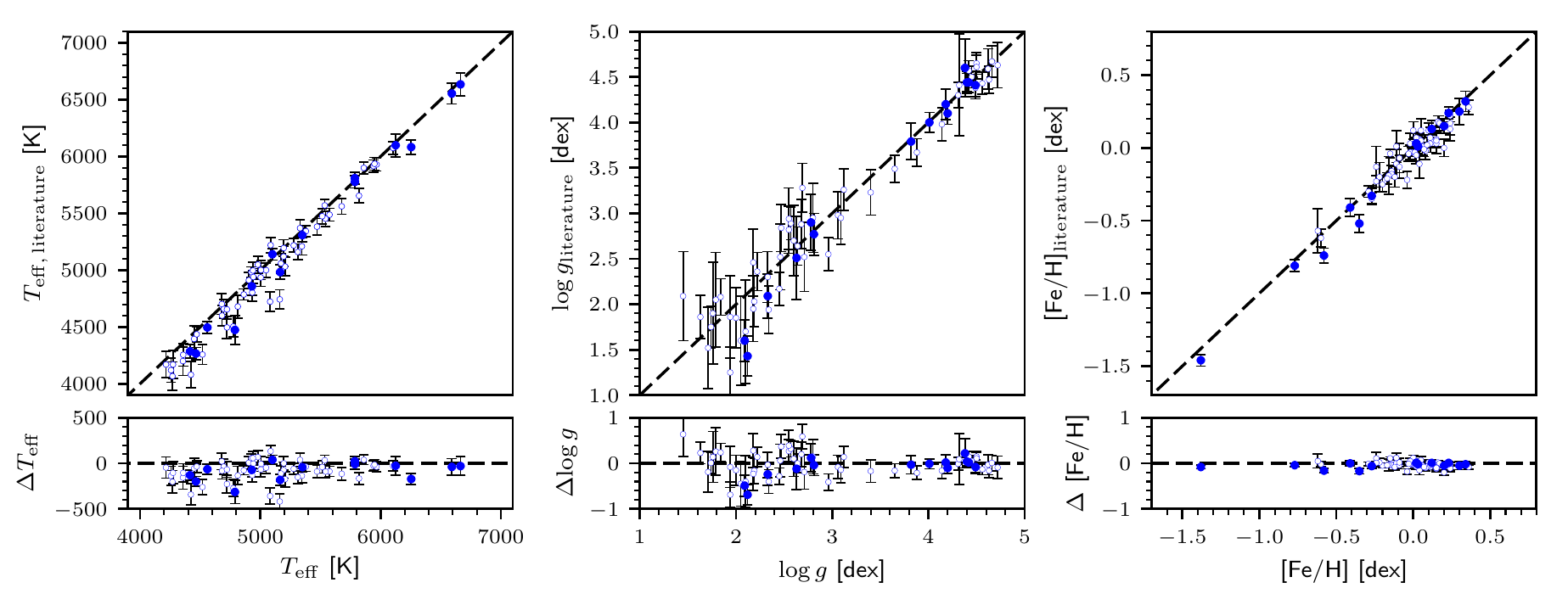}
\caption{\label{fig:Teff_logg_FeH_all_vis} Same as Fig.~\ref{fig:Teff_logg_FeH_all} but restricting the analysis to the \ion{Fe}{i} and \ion{Fe}{ii} lines found in the optical wavelength region covered by the VIS channel of CARMENES.}
\end{figure*}

\begin{figure}
\centering
\includegraphics{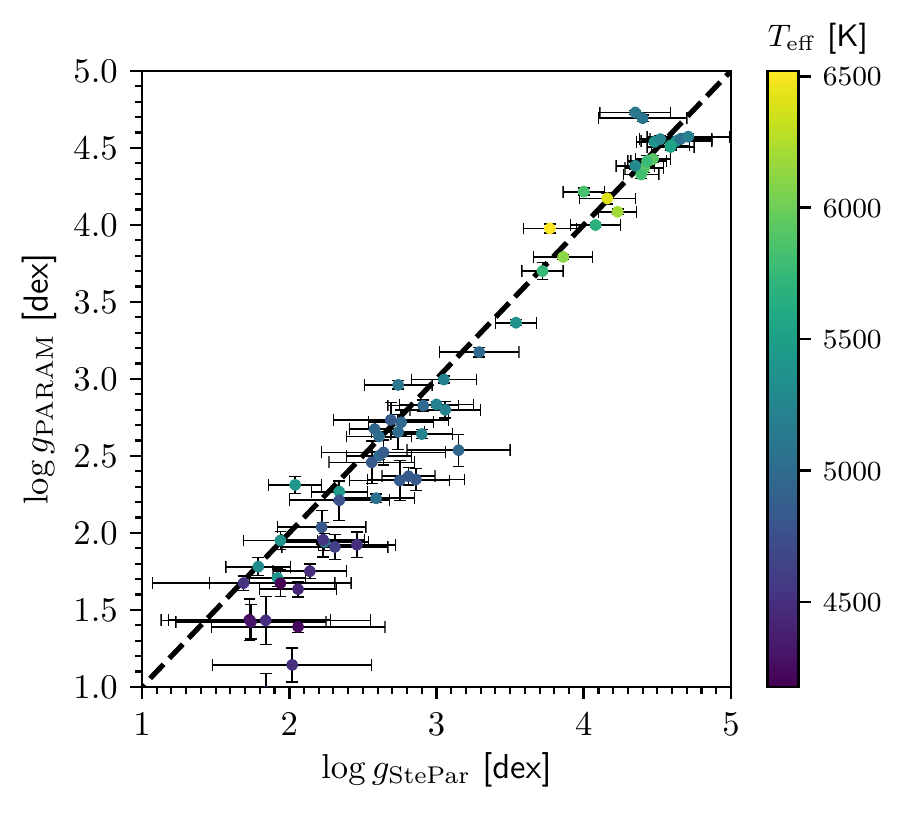}
\caption{\label{fig:loggdr2} Surface gravities, $\log{g}$, derived for the sample with {\sc StePar} versus those obtained with the code {\sc PARAM}, adopting the distances from {\it Gaia} DR2.}
\end{figure}

\begin{figure}
\centering
\includegraphics{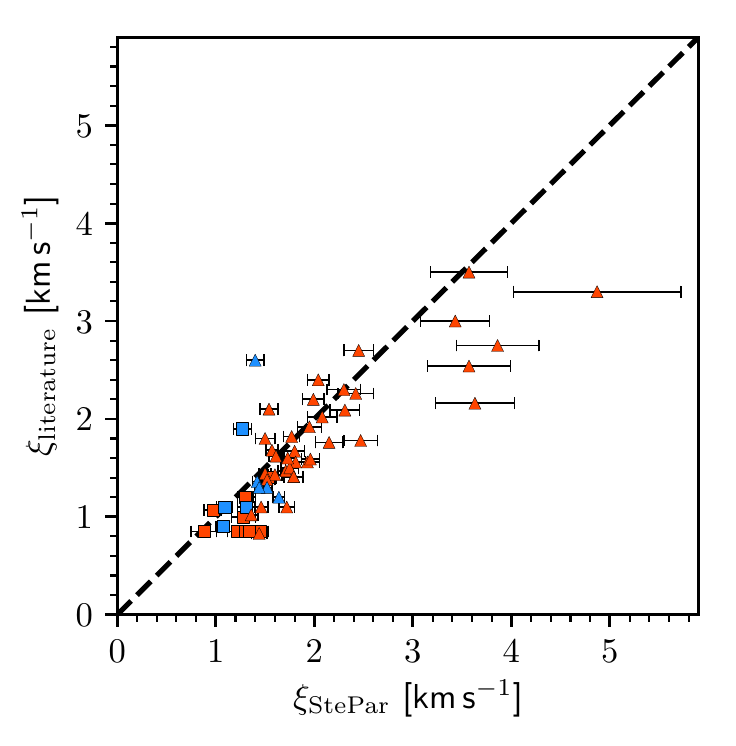}
\caption{\label{fig:vmicro} Micro-turbulent velocity derived for the sample with {\sc StePar}, $\xi_{\rm StePar}$, versus literature values. Symbols are the same as in Fig.~\ref{fig:sample_histogram}.}
\end{figure}

\begin{figure}
\centering
\includegraphics{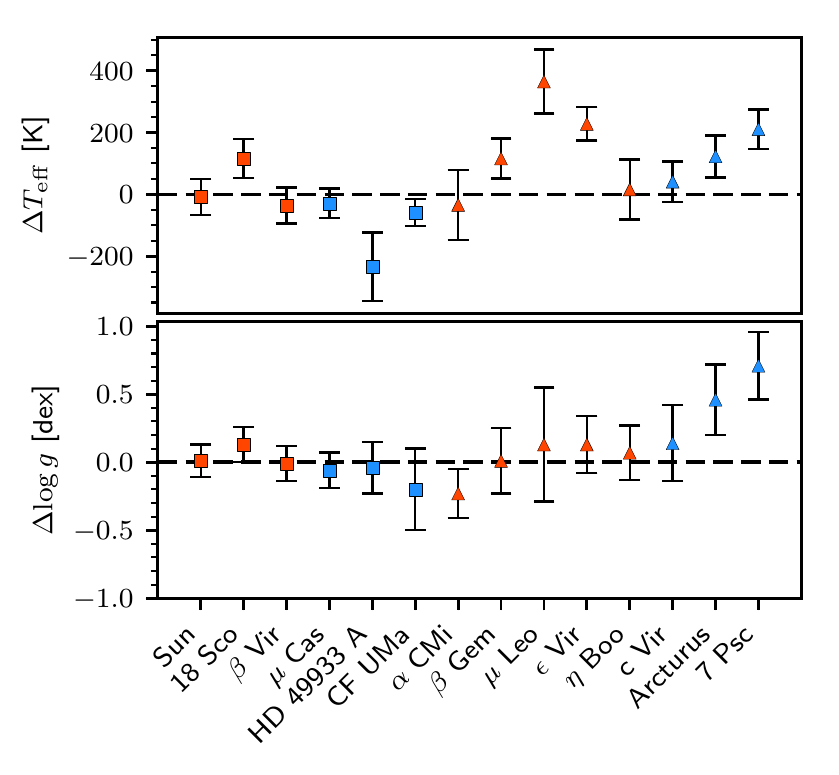}
\caption{\label{fig:benchmarks_all} Differences in $T_{\rm eff}$ and $\log{g}$ between this work and \citet{Hei15}, with updated values from \citet{Jof18}, for the \emph{Gaia} benchmark stars in our sample. Symbols are the same as in Fig.~\ref{fig:sample_histogram}.}
\end{figure}

\begin{figure}
\centering
\includegraphics[width=0.49\textwidth]{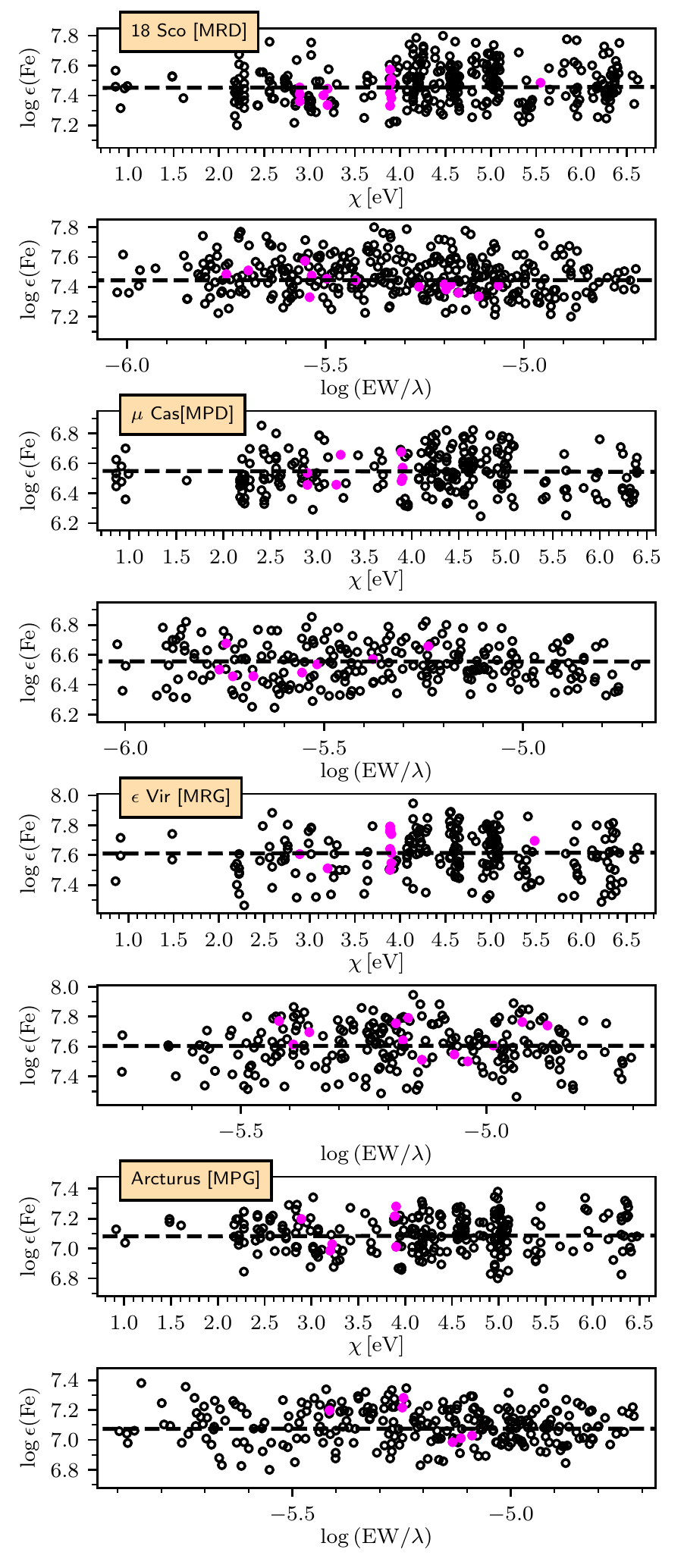}
\caption{\label{fig:abun} {\it From top to bottom:} line iron abundance retrieved by {\sc StePar} for the final solution of the four reference stars: 18~Sco, $\mu$~Cas, $\epsilon$~Vir, and Arcturus. $\log{\epsilon}$(Fe~{\sc i}) stands for the Fe abundance returned by the Fe lines, while $\log{({\rm EW}/\lambda)}$ is their reduced EWs. Unfilled black dots represent Fe~{\sc i} lines, whereas pink dots are Fe~{\sc ii} lines. The dashed black lines represent the least-squares fit to the data points.}
\end{figure}

In Table~\ref{tab:par_stars_stepar} we give the stellar atmospheric parameters of the sample computed with {\sc StePar}. These were obtained after matching the corresponding \ion{Fe}{i} and \ion{Fe}{ii} line lists to the stars according to their reference parameters reported in Table~\ref{tab:par_stars_ref}.

We also performed the analysis of the sample with the $EW$ method taking into account only the \ion{Fe}{i} and \ion{Fe}{ii} lines found in the optical region covered by the VIS channel of the CARMENES instrument. The parameters thus obtained can be found in Table~\ref{tab:par_stars_stepar_vis}. Unfortunately, we could not attempt to analyse the NIR in the same manner because of the scarcity of \ion{Fe}{ii} lines above 9\,600~{\AA}.

In Fig.~\ref{fig:kiel_diagram} we display a Kiel diagram, i.e. $\log{g}$ versus $\log{T_{\rm eff}}$, of our sample as computed with {\sc StePar}, along with the Yale-Potsdam Stellar Isochrones \citep[YaPSI,][]{Spa17} at solar metallicity, namely ${\rm Z}=0.016$. Overall, we found no disparity between our derived values and the region of the parameter space covered by the isochrones. As pointed out by \citet{Tab19}, {\sc StePar} returns slightly higher effective temperatures for F-type dwarfs. Five luminous, G-type, giant stars ($\beta$~Dra, F~Hya, $\epsilon$~Leo, 37~LMi, and $\zeta$~Mon) are located at an anomalous position in the Kiel diagram. According to \citet{Luc14}, these stars are thought to be the evolved counterparts of early F- to B-type main-sequence stars that have reached the He-burning evolutionary stage.

In the cool regime, i.e. K-type stars, where stellar spectra become increasingly more crowded, the continuum placement is more uncertain, and the iron lines are subject to blending with other spectral features. On the other hand, sufficiently strong iron lines become increasingly scarce towards early F-type stars. This has a strong impact on the computed errors in the stellar atmospheric parameters, in particular the effective temperature, and the line-to-line scatter in [Fe/H], as shown in Figs.~\ref{fig:T_eff_errors_all} and \ref{fig:fe_dispersion}, respectively.

In Figs.~\ref{fig:Teff_logg_FeH_all} and \ref{fig:Teff_logg_FeH_all_vis} we compare the stellar atmospheric parameters computed with {\sc StePar} with values from the literature \citep{McW90, Hei03, All04, Val05, Hek07, Liu07, Sou08, Tak08, Lyu10, Wu11, Thy12, San13, Mor14, Luc14, Jof14, Jof15, Sil15, Jof18}, taking into account the VIS and NIR channels simultaneously, and only the VIS channel, respectively. To explore possible sources of potential systematic trends or offsets, we followed the Monte Carlo method implemented in \citet{Tab18}. We generated 10\,000 synthetic samples based on our derived stellar atmospheric parameters. We computed all data points in each of these artificial samples by means of a normal distribution centred at the original measurements, and took the uncertainties in each parameter as the width of the distribution. The summary of the Monte Carlo simulations can be found in Table~\ref{tab:mc}. We computed the Pearson and Spearman correlation coefficients, which quantify the degree of correlation between any two given variables. We found a significant correlation in the differences  between our own $T_{\rm eff}$ values and the literature versus the literature values. However, no such correlation was found in the derived $\log{g}$ and [Fe/H] values.

At first glance, it seems that our temperature scale has an intrinsic systematic error with respect to the literature values. The offset appears to be linked to the fact that we now include the NIR channel, given that the correlation diminishes when we restrict the analysis to the iron lines found in the VIS channel. Although the {\sc StePar} code could be thought to be the underlying reason for this correlation, we are not comparing the same temperature scale. In other words, we now take into account iron absorption lines in a wavelength region that is different from most studies found in the literature. In addition, this offset is more noticeable for the coolest stars. The former result could arise from the fact that the NIR lines are more sensitive to the effective temperature than the optical lines, at least for the cool stars. In other words, although the inclusion of the NIR in the analysis does not bring extreme differences of the derived stellar parameters with respect to the analysis using the optical range, it seems to reveal a deeper $T_{\rm eff}$ scale as suggested by the meaningful correlation found in Table~\ref{tab:mc} as well as Figs.~\ref{fig:Teff_logg_FeH_all} and \ref{fig:Teff_logg_FeH_all_vis}. 

In Fig.~\ref{fig:loggdr2} we show the values of $\log{g}$ derived with {\sc StePar} against those obtained adopting the distances from {\it Gaia} DR2 \citep{gaia}, if available, and the Hipparcos mission \citep{Lee07}. We computed the latter $\log{g}$ values by means of the {\sc PARAM} web interface\footnote{\tt http://stev.oapd.inaf.it/cgi-bin/param} \citep{Sil06, Rod14, Rod17}, which employs a Bayesian approach to derive the stellar parameters, including stellar age, mass and radius. The $\log{g}$ values obtained with PARAM can be found in Tables \ref{tab:par_stars_stepar} and \ref{tab:par_stars_stepar_vis}. Following the Monte Carlo method described above, we found a systematic offset of 0.15$\pm$0.38~dex. The Pearson and Spearman correlation coefficients, which are $r_p=-0.302\pm0.093$ and $r_s=0.259\pm0.104$, respectively, reveal a correlation of around 9\%, which is slightly lower than previous works \citep[see e.g.][]{Tab17}.

Regarding the micro-turbulent velocity, Fig.~\ref{fig:vmicro} shows the values of $\xi$ obtained with {\sc StePar} against the literature. Our derived values for $\xi$ are compatible to the literature values to a large extent. However, six stars (i.e. $\beta$~Dra, F~Hya, $\zeta$~Mon, $\sigma$~Oph, $\theta$~Her, and HD~77912), with computed $\xi$ values larger than 3~km/s, show larger deviations with respect to the literature, which can be as large as 1.6 km\,s$^{-1}$. as in the case of the star $\zeta$~Mon. In addition, we retrieved a significantly lower $\xi$ value for the star $\upsilon$~Boo compared to the literature. Although $\xi$ and [Fe/H] are thought to be partially degenerate \citep{Val05}, we fail to identify the impact that such high or low $\xi$ values have on [Fe/H] for these stars in our analysis. For example, a difference of 1.6 km\,s$^{-1}$ in $\xi$ for the star $\zeta$~Mon leads to a difference of only 0.07~dex in [Fe/H] between the literature and the analysis with {\sc StePar}, and both computed and literature values are compatible within error bars.

A closer look at the comparison between our parameter determinations and the \emph{Gaia} benchmark star parameters from \citet{Hei15}, with updated values from \citet{Jof18}, can be found in Fig.~\ref{fig:benchmarks_all}. We find good agreement between our derived values and the fundamental $T_{\rm eff}$ and $\log{g}$, i.e. derived from the fundamental relations $L=4\pi R^2\sigma T_{\rm eff}^4$ and $g=GM/R^2$, respectively, by means of specific information that is available for these stars, such as the parallax, the angular diameter, and the bolometric flux. Nonetheless, we note four outliers in $T_{\rm eff}$ ($\Delta T_{\rm eff} > 200$\,K) and two in $\log{g}$ ($\Delta\log{g} > 0.25$\,dex). Among the outliers in $\log{g}$ are Arcturus and 7~Psc. According to \citet{Hei15}, the $\log{g}$ value of Arcturus remains uncertain, with literature values ranging from 1.4 up to 2.0\,dex, while both the $T_{\rm eff}$ and $\log{g}$ values for the star 7~Psc are, in fact, not recommended for use as reference values. Among the outliers in $T_{\rm eff}$ are the stars HD~49933, $\mu$~Leo, $\epsilon$~Vir, and 7~Psc. As stated by \citet{Hei15}, the fundamental $T_{\rm eff}$ value for the stars $\epsilon$~Vir and $\mu$~Leo is significantly lower ($\sim$ 3\%) than the value derived in spectroscopic studies. Lastly, at the hot regime, the typical spectroscopic $T_{\rm eff}$ values computed for the star HD~49933 are generally larger.

Lastly, in Fig.~\ref{fig:abun} we show the final Fe~{\sc i} and Fe~{\sc ii} abundances versus the excitation potential and the reduced equivalent width of the lines, for the four reference CARMENES spectra (18~Sco, $\mu$~Cas, $\epsilon$~Vir, and Arcturus).

\section{Conclusions}
\label{sec:conclusions}

In this work, we have expanded previous optical \ion{Fe}{i} and \ion{Fe}{ii} line lists into the wavelength range covered by CARMENES, i.e. from 5\,300 to 17\,100\,{\AA}. The line lists are suited for FGK-type stars and relate to metal-rich dwarfs (MRD), metal-poor dwarfs (MPD), metal-rich giants (MRG), and metal-poor giants (MPG). For the first time, we provide \ion{Fe}{i} and \ion{Fe}{ii} lines in the wavelength region between 6\,800\,{\AA} and 10\,000\,{\AA}. Altogether, these new line lists contain 653 \ion{Fe}{i} and 23 \ion{Fe}{ii} lines, of which 351 and 8 are new additions to the line lists compiled in \citet{Tab19}, respectively. This implies more than doubling the number of \ion{Fe}{i} and \ion{Fe}{ii} lines useful for abundance and radial-velocity analyses. The availability of these \ion{Fe}{i} and \ion{Fe}{ii} line lists is also an asset for other new high-resolution near-infrared spectrographs such as SPIRou, GIANO, CRIRES+, IRD, HPF, and NIRPS that also provide wavelength coverage in the near-infrared wavelength region.

We have reported that the star c~Gem (HD~62285) is a new SB2 system, as shown by the cross-correlation with an atlas spectrum of Arcturus.

In addition, we have computed an homogenised set of stellar atmospheric parameters for a sample of 65 FGK-type stars observed with CARMENES by means of the $EW$ method. We made a comprehensive comparison of our $T_{\rm eff}$, $\log{g}$, and [Fe/H] values with those of virtually all relevant determinations of stellar atmospheric parameters of FGK-type stars. Our parameter determinations are in good agreement with the literature values in general, particularly with the region of the parameter space covered by the YaPSI isochrones \citep{Spa17} and the \emph{Gaia} benchmark stars \citep{Jof14, Jof18, Hei15}. The scarcity of \ion{Fe}{ii} lines in the NIR wavelength range covered by CARMENES prevented us from performing the stellar parameter determinations using this spectral region alone. However, when using both VIS and NIR CARMENES channel data, we found a broader $T_{\rm eff}$ scale that seems to be linked to a higher sensitivity to effective temperature of the iron lines found in the NIR region.

The line selections provided in this work will be useful for the spectroscopic analysis of any FGK-type star simultaneously observed in the optical and near-infrared wavelength regions. Finally, in a forthcoming publication we plan to expand optical line lists of additional chemical species into the NIR covered by CARMENES and thus assess the impact of the near-infrared wavelength region upon chemical abundance computations for FGK-type stars.

\section*{Acknowledgements}

CARMENES is an instrument for the Centro Astron\'{o}mico Hispano en Andaluc\'{i}a at Calar Alto (CAHA). CARMENES is funded by the German Max-Plank Gesellschaft (MPG), the Spanish Consejo Superior de Investigaciones Cient\'{i}ficas (CSIC), the European Union through FEDER/ERF FICTS-2011-02 funds, and the members of the CARMENES Consortium (Max-Plank-Institut f\"{u}r Astronomie, Instituto de Astrof\'{i}sica de Andaluc\'{i}a, Landessternwarte K\"{o}nigstuhl, Institut de Ci\`{e}ncies de l'Espai, Institut f\"{u}r Astrophysik G\"{o}ttingen, Universidad Complutense de Madrid, Th\"{u}ringer Landessternwarte Tautenberg, Instituto de Astrof\'{i}sica de Canarias, Hamburger Sternwarte, Centro de Astrobiolog\'{i}a and Centro Astron\'{o}mico Hispano en Andaluc\'{i}a), with additional contributions by the Spanish Ministerio de Econom\'{i}a y Empresa, the German Science Foundation through the Major Research Instrumentation Programme and DFG Research Unit FOR2544 "Blue Planets around Red Stars", the Klaus Tschira Stiftung, the states of Baden-W\"{u}rttemberg and Niedersachsen, and by the Junta de Andaluc\'{i}a. The authors acknowledge financial support from the Funda\c c\~ao para a Ci\^encia e a Tecnologia (FCT) through national funds (PTDC/FIS-AST/28953/2017) and by Fundo Europeu de Desenvolvimento Regional (FEDER) through COMPETE2020 - Programa Operacional Competitividade e Internacionaliza\c c\~ao (POCI-01-0145-FEDER-028953), and the Spanish {Ministerio de Ciencia, Innovaci\'{o}n y Universidades}, {the Universidad Complutense de Madrid}, and the Fondo Europeo de Desarrollo Regional (FEDER/ERF) through fellowship FPU15/01476, and projects AYA2016-79425-C3-1/2/3-P. J.I.G.H. acknowledges financial support from the Spanish Ministerio de Ciencia, Innovaci\'{o}n y Universidades under the 2013 Ram\'{o}n y Cajal programme RYC-2013-14875, and from the project AYA2017-86389-P. This work has made use of the VALD database, operated at Uppsala University, the Institute of Astronomy RAS in Moscow, and the University of Vienna. We thank Calar Alto Observatory for the allocation of director's discretionary time to this programme. E.~M. would also like to warmly thank the staff at the Hamburger Sternwarte for their hospitality during his stay funded by project EST18/00162. Based on data from the CARMENES data archive at CAB (INTA-CSIC).




\bibliographystyle{mnras}
\bibliography{pa} 

\begin{thebibliography}{}
\makeatletter
\relax
\def\mn@urlcharsother{\let\do\@makeother \do\$\do\&\do\#\do\^\do\_\do\%\do\~}
\def\mn@doi{\begingroup\mn@urlcharsother \@ifnextchar [ {\mn@doi@}
  {\mn@doi@[]}}
\def\mn@doi@[#1]#2{\def\@tempa{#1}\ifx\@tempa\@empty \href
  {http://dx.doi.org/#2} {doi:#2}\else \href {http://dx.doi.org/#2} {#1}\fi
  \endgroup}
\def\mn@eprint#1#2{\mn@eprint@#1:#2::\@nil}
\def\mn@eprint@arXiv#1{\href {http://arxiv.org/abs/#1} {{\tt arXiv:#1}}}
\def\mn@eprint@dblp#1{\href {http://dblp.uni-trier.de/rec/bibtex/#1.xml}
  {dblp:#1}}
\def\mn@eprint@#1:#2:#3:#4\@nil{\def\@tempa {#1}\def\@tempb {#2}\def\@tempc
  {#3}\ifx \@tempc \@empty \let \@tempc \@tempb \let \@tempb \@tempa \fi \ifx
  \@tempb \@empty \def\@tempb {arXiv}\fi \@ifundefined
  {mn@eprint@\@tempb}{\@tempb:\@tempc}{\expandafter \expandafter \csname
  mn@eprint@\@tempb\endcsname \expandafter{\@tempc}}}

\bibitem[\protect\citeauthoryear{{Adibekyan}, {Gonz{\'a}lez Hern{\'a}ndez},
  {Delgado Mena}, {Sousa}, {Santos}, {Israelian}, {Figueira}  \& {Bertran de
  Lis}}{{Adibekyan} et~al.}{2014}]{Adi14}
{Adibekyan} V.~Z.,  {Gonz{\'a}lez Hern{\'a}ndez} J.~I.,  {Delgado Mena} E.,
  {Sousa} S.~G.,  {Santos} N.~C.,  {Israelian} G.,  {Figueira} P.,   {Bertran
  de Lis} S.,  2014, \mn@doi [\aap] {10.1051/0004-6361/201423435}, \href
  {http://adsabs.harvard.edu/abs/2014A%26A...564L..15A} {564, L15}

\bibitem[\protect\citeauthoryear{{Allende Prieto}, {Barklem}, {Lambert}  \&
  {Cunha}}{{Allende Prieto} et~al.}{2004}]{All04}
{Allende Prieto} C.,  {Barklem} P.~S.,  {Lambert} D.~L.,   {Cunha} K.,  2004,
  \mn@doi [\aap] {10.1051/0004-6361:20035801}, \href
  {http://adsabs.harvard.edu/abs/2004A%26A...420..183A} {420, 183}

\bibitem[\protect\citeauthoryear{{Allende Prieto} et~al.,}{{Allende Prieto}
  et~al.}{2008}]{All08}
{Allende Prieto} C.,  et~al., 2008, \mn@doi [Astronomische Nachrichten]
  {10.1002/asna.200811080}, \href
  {http://adsabs.harvard.edu/abs/2008AN....329.1018A} {329, 1018}

\bibitem[\protect\citeauthoryear{{Andreasen}, {Sousa}, {Delgado Mena},
  {Santos}, {Tsantaki}, {Rojas-Ayala}  \& {Neves}}{{Andreasen}
  et~al.}{2016}]{And16}
{Andreasen} D.~T.,  {Sousa} S.~G.,  {Delgado Mena} E.,  {Santos} N.~C.,
  {Tsantaki} M.,  {Rojas-Ayala} B.,   {Neves} V.,  2016, \mn@doi [\aap]
  {10.1051/0004-6361/201527308}, \href
  {http://adsabs.harvard.edu/abs/2016A%26A...585A.143A} {585, A143}

\bibitem[\protect\citeauthoryear{{Artigau} et~al.,}{{Artigau}
  et~al.}{2014}]{Art14}
{Artigau} {\'E}.,  et~al., 2014, in Ground-based and Airborne Instrumentation
  for Astronomy V. p. 914715, \mn@doi{10.1117/12.2055663}

\bibitem[\protect\citeauthoryear{{Auri{\`e}re}}{{Auri{\`e}re}}{2003}]{Aur03}
{Auri{\`e}re} M.,  2003, in {Arnaud} J.,  {Meunier} N.,  eds,  EAS Publications
  Series Vol. 9, EAS Publications Series. p.~105

\bibitem[\protect\citeauthoryear{{Bensby}, {Feltzing}  \& {Oey}}{{Bensby}
  et~al.}{2014}]{Ben14}
{Bensby} T.,  {Feltzing} S.,   {Oey} M.~S.,  2014, \mn@doi [\aap]
  {10.1051/0004-6361/201322631}, \href
  {http://adsabs.harvard.edu/abs/2014A%26A...562A..71B} {562, A71}

\bibitem[\protect\citeauthoryear{{Blanco-Cuaresma}}{{Blanco-Cuaresma}}{2019}]{Bla19}
{Blanco-Cuaresma} S.,  2019, \mn@doi [\mnras] {10.1093/mnras/stz549}, \href
  {https://ui.adsabs.harvard.edu/abs/2019MNRAS.486.2075B} {486, 2075}

\bibitem[\protect\citeauthoryear{{Blanco-Cuaresma}, {Soubiran}, {Heiter}  \&
  {Jofr{\'e}}}{{Blanco-Cuaresma} et~al.}{2014}]{Bla14}
{Blanco-Cuaresma} S.,  {Soubiran} C.,  {Heiter} U.,   {Jofr{\'e}} P.,  2014,
  {iSpec: Stellar atmospheric parameters and chemical abundances}, Astrophysics
  Source Code Library (\mn@eprint {ascl} {1409.006})

\bibitem[\protect\citeauthoryear{{Caballero} et~al.,}{{Caballero}
  et~al.}{2016}]{Cab16}
{Caballero} J.~A.,  et~al., 2016, in \procspie. p. 99100E,
  \mn@doi{10.1117/12.2233574}

\bibitem[\protect\citeauthoryear{{Clough}, {Shephard}, {Mlawer}, {Delamere},
  {Iacono}, {Cady-Pereira}, {Boukabara}  \& {Brown}}{{Clough}
  et~al.}{2005}]{Clo05}
{Clough} S.~A.,  {Shephard} M.~W.,  {Mlawer} E.~J.,  {Delamere} J.~S.,
  {Iacono} M.~J.,  {Cady-Pereira} K.,  {Boukabara} S.,   {Brown} P.~D.,  2005,
  \mn@doi [\jqsrt] {10.1016/j.jqsrt.2004.05.058}, \href
  {https://ui.adsabs.harvard.edu/abs/2005JQSRT..91..233C} {91, 233}

\bibitem[\protect\citeauthoryear{{De Medeiros}, {do Nascimento},
  {Sankarankutty}, {Costa}  \& {Maia}}{{De Medeiros} et~al.}{2000}]{Med00}
{De Medeiros} J.~R.,  {do Nascimento} J.~D. J.,  {Sankarankutty} S.,  {Costa}
  J.~M.,   {Maia} M.~R.~G.,  2000, \aap, \href
  {https://ui.adsabs.harvard.edu/abs/2000A&A...363..239D} {363, 239}

\bibitem[\protect\citeauthoryear{{De Medeiros}, {Udry}, {Burki}  \&
  {Mayor}}{{De Medeiros} et~al.}{2002}]{Med02}
{De Medeiros} J.~R.,  {Udry} S.,  {Burki} G.,   {Mayor} M.,  2002, \mn@doi
  [\aap] {10.1051/0004-6361:20021214}, \href
  {https://ui.adsabs.harvard.edu/abs/2002A&A...395...97D} {395, 97}

\bibitem[\protect\citeauthoryear{{De Silva} et~al.,}{{De Silva}
  et~al.}{2015}]{DeS15}
{De Silva} G.~M.,  et~al., 2015, \mn@doi [\mnras] {10.1093/mnras/stv327}, \href
  {http://adsabs.harvard.edu/abs/2015MNRAS.449.2604D} {449, 2604}

\bibitem[\protect\citeauthoryear{{Delgado Mena} et~al.,}{{Delgado Mena}
  et~al.}{2018}]{Del18}
{Delgado Mena} E.,  et~al., 2018, \mn@doi [\aap] {10.1051/0004-6361/201833152},
  \href {http://adsabs.harvard.edu/abs/2018A%26A...619A...2D} {619, A2}

\bibitem[\protect\citeauthoryear{{Famaey}, {Jorissen}, {Luri}, {Mayor}, {Udry},
  {Dejonghe}  \& {Turon}}{{Famaey} et~al.}{2005}]{Fam05}
{Famaey} B.,  {Jorissen} A.,  {Luri} X.,  {Mayor} M.,  {Udry} S.,  {Dejonghe}
  H.,   {Turon} C.,  2005, \mn@doi [\aap] {10.1051/0004-6361:20041272}, \href
  {http://adsabs.harvard.edu/abs/2005A%26A...430..165F} {430, 165}

\bibitem[\protect\citeauthoryear{{Gaia Collaboration} et~al.,}{{Gaia
  Collaboration} et~al.}{2018}]{gaia}
{Gaia Collaboration} et~al., 2018, \mn@doi [\aap]
  {10.1051/0004-6361/201833051}, \href
  {https://ui.adsabs.harvard.edu/abs/2018A&A...616A...1G} {616, A1}

\bibitem[\protect\citeauthoryear{{Gilmore} et~al.,}{{Gilmore}
  et~al.}{2012}]{Gil12}
{Gilmore} G.,  et~al., 2012, The Messenger, \href
  {http://adsabs.harvard.edu/abs/2012Msngr.147...25G} {147, 25}

\bibitem[\protect\citeauthoryear{{Girard} et~al.,}{{Girard}
  et~al.}{2000}]{Gira00}
{Girard} T.~M.,  et~al., 2000, \mn@doi [\aj] {10.1086/301353}, \href
  {https://ui.adsabs.harvard.edu/abs/2000AJ....119.2428G} {119, 2428}

\bibitem[\protect\citeauthoryear{{Gontcharov}}{{Gontcharov}}{2006}]{Gon06}
{Gontcharov} G.~A.,  2006, \mn@doi [Astronomy Letters]
  {10.1134/S1063773706110065}, \href
  {http://adsabs.harvard.edu/abs/2006AstL...32..759G} {32, 759}

\bibitem[\protect\citeauthoryear{{Gordon}, {Rothman}, {Tan}, {Kochanov}  \&
  {Hill}}{{Gordon} et~al.}{2017}]{Gor17}
{Gordon} I.~E.,  {Rothman} L.~S.,  {Tan} Y.,  {Kochanov} R.~V.,   {Hill} C.,
  2017, in 72nd International Symposium on Molecular Spectroscopy. p.~TJ08,
  \mn@doi{10.15278/isms.2017.TJ08}

\bibitem[\protect\citeauthoryear{{Gray}}{{Gray}}{2018}]{Gra18}
{Gray} D.~F.,  2018, \mn@doi [\apj] {10.3847/1538-4357/aae9e6}, \href
  {https://ui.adsabs.harvard.edu/abs/2018ApJ...869...81G} {869, 81}

\bibitem[\protect\citeauthoryear{{Gray} \& {Toner}}{{Gray} \&
  {Toner}}{1986}]{Gra86}
{Gray} D.~F.,  {Toner} C.~G.,  1986, \mn@doi [\apj] {10.1086/164681}, \href
  {https://ui.adsabs.harvard.edu/abs/1986ApJ...310..277G} {310, 277}

\bibitem[\protect\citeauthoryear{{Gustafsson}, {Edvardsson}, {Eriksson},
  {J{\o}rgensen}, {Nordlund}  \& {Plez}}{{Gustafsson} et~al.}{2008}]{Gus08}
{Gustafsson} B.,  {Edvardsson} B.,  {Eriksson} K.,  {J{\o}rgensen} U.~G.,
  {Nordlund} {\AA}.,   {Plez} B.,  2008, \mn@doi [\aap]
  {10.1051/0004-6361:200809724}, \href
  {http://adsabs.harvard.edu/abs/2008A%26A...486..951G} {486, 951}

\bibitem[\protect\citeauthoryear{{Hatzes} \& {CRIRES+ Team}}{{Hatzes} \&
  {CRIRES+ Team}}{2017}]{Hat17}
{Hatzes} A.,  {CRIRES+ Team} 2017, in American Astronomical Society Meeting
  Abstracts \#230. p. 117.02

\bibitem[\protect\citeauthoryear{{Hayes} et~al.,}{{Hayes} et~al.}{2018}]{Hay18}
{Hayes} C.~R.,  et~al., 2018, \mn@doi [\apj] {10.3847/1538-4357/aa9cec}, \href
  {https://ui.adsabs.harvard.edu/abs/2018ApJ...852...49H} {852, 49}

\bibitem[\protect\citeauthoryear{{Heiter} \& {Luck}}{{Heiter} \&
  {Luck}}{2003}]{Hei03}
{Heiter} U.,  {Luck} R.~E.,  2003, \mn@doi [\aj] {10.1086/378366}, \href
  {http://adsabs.harvard.edu/abs/2003AJ....126.2015H} {126, 2015}

\bibitem[\protect\citeauthoryear{{Heiter}, {Jofr{\'e}}, {Gustafsson}, {Korn},
  {Soubiran}  \& {Th{\'e}venin}}{{Heiter} et~al.}{2015}]{Hei15}
{Heiter} U.,  {Jofr{\'e}} P.,  {Gustafsson} B.,  {Korn} A.~J.,  {Soubiran} C.,
   {Th{\'e}venin} F.,  2015, \mn@doi [\aap] {10.1051/0004-6361/201526319},
  \href {http://adsabs.harvard.edu/abs/2015A%26A...582A..49H} {582, A49}

\bibitem[\protect\citeauthoryear{{Hekker} \& {Mel{\'e}ndez}}{{Hekker} \&
  {Mel{\'e}ndez}}{2007}]{Hek07}
{Hekker} S.,  {Mel{\'e}ndez} J.,  2007, \mn@doi [\aap]
  {10.1051/0004-6361:20078233}, \href
  {http://adsabs.harvard.edu/abs/2007A%26A...475.1003H} {475, 1003}

\bibitem[\protect\citeauthoryear{{Hinkle}, {Wallace}, {Valenti}  \&
  {Harmer}}{{Hinkle} et~al.}{2000}]{Hin00}
{Hinkle} K.,  {Wallace} L.,  {Valenti} J.,   {Harmer} D.,  2000, {Visible and
  Near Infrared Atlas of the Arcturus Spectrum 3727-9300 A}.
Astronomical Society of the Pacific

\bibitem[\protect\citeauthoryear{{Jofr{\'e}} et~al.,}{{Jofr{\'e}}
  et~al.}{2014}]{Jof14}
{Jofr{\'e}} P.,  et~al., 2014, \mn@doi [\aap] {10.1051/0004-6361/201322440},
  \href {http://adsabs.harvard.edu/abs/2014A%26A...564A.133J} {564, A133}

\bibitem[\protect\citeauthoryear{{Jofr{\'e}}, {Petrucci}, {Saffe}, {Saker}, {de
  la Villarmois}, {Chavero}, {G{\'o}mez}  \& {Mauas}}{{Jofr{\'e}}
  et~al.}{2015}]{Jof15}
{Jofr{\'e}} E.,  {Petrucci} R.,  {Saffe} C.,  {Saker} L.,  {de la Villarmois}
  E.~A.,  {Chavero} C.,  {G{\'o}mez} M.,   {Mauas} P.~J.~D.,  2015, \mn@doi
  [\aap] {10.1051/0004-6361/201424474}, \href
  {http://adsabs.harvard.edu/abs/2015A%26A...574A..50J} {574, A50}

\bibitem[\protect\citeauthoryear{{Jofr{\'e}}, {Heiter}, {Tucci Maia},
  {Soubiran}, {Worley}, {Hawkins}, {Blanco-Cuaresma}  \& {Rodrigo}}{{Jofr{\'e}}
  et~al.}{2018}]{Jof18}
{Jofr{\'e}} P.,  {Heiter} U.,  {Tucci Maia} M.,  {Soubiran} C.,  {Worley}
  C.~C.,  {Hawkins} K.,  {Blanco-Cuaresma} S.,   {Rodrigo} C.,  2018, \mn@doi
  [Research Notes of the American Astronomical Society]
  {10.3847/2515-5172/aadc61}, \href
  {http://adsabs.harvard.edu/abs/2018RNAAS...2c.152J} {2, 152}

\bibitem[\protect\citeauthoryear{Jofr{\'e}, {Heiter}  \& Soubiran}{Jofr{\'e}
  et~al.}{2019}]{Jof19}
Jofr{\'e} P.,  {Heiter} U.,   Soubiran C.,  2019, \mn@doi [\araa]
  {10.1146/annurev-astro-091918-104509}, 57, 5

\bibitem[\protect\citeauthoryear{{Karata{\c s}}, {Bilir}, {Eker}  \&
  {Demircan}}{{Karata{\c s}} et~al.}{2004}]{Kar04}
{Karata{\c s}} Y.,  {Bilir} S.,  {Eker} Z.,   {Demircan} O.,  2004, \mn@doi
  [\mnras] {10.1111/j.1365-2966.2004.07588.x}, \href
  {http://adsabs.harvard.edu/abs/2004MNRAS.349.1069K} {349, 1069}

\bibitem[\protect\citeauthoryear{{Kasting}, {Whitmire}  \&
  {Reynolds}}{{Kasting} et~al.}{1993}]{Kas93}
{Kasting} J.~F.,  {Whitmire} D.~P.,   {Reynolds} R.~T.,  1993, \mn@doi
  [\icarus] {10.1006/icar.1993.1010}, \href
  {https://ui.adsabs.harvard.edu/abs/1993Icar..101..108K} {101, 108}

\bibitem[\protect\citeauthoryear{{Kausch} et~al.,}{{Kausch}
  et~al.}{2015}]{Kau15}
{Kausch} W.,  et~al., 2015, \mn@doi [\aap] {10.1051/0004-6361/201423909}, \href
  {https://ui.adsabs.harvard.edu/abs/2015A&A...576A..78K} {576, A78}

\bibitem[\protect\citeauthoryear{{Keenan} \& {McNeil}}{{Keenan} \&
  {McNeil}}{1989}]{Kee89}
{Keenan} P.~C.,  {McNeil} R.~C.,  1989, \mn@doi [\apjs] {10.1086/191373}, \href
  {https://ui.adsabs.harvard.edu/abs/1989ApJS...71..245K} {71, 245}

\bibitem[\protect\citeauthoryear{{Kopparapu} et~al.,}{{Kopparapu}
  et~al.}{2013}]{Kop13}
{Kopparapu} R.~K.,  et~al., 2013, \mn@doi [\apj] {10.1088/0004-637X/765/2/131},
  \href {https://ui.adsabs.harvard.edu/abs/2013ApJ...765..131K} {765, 131}

\bibitem[\protect\citeauthoryear{{Kotani} et~al.,}{{Kotani}
  et~al.}{2014}]{Kot14}
{Kotani} T.,  et~al., 2014, in Ground-based and Airborne Instrumentation for
  Astronomy V. p. 914714, \mn@doi{10.1117/12.2055075}

\bibitem[\protect\citeauthoryear{{Kupka}, {Piskunov}, {Ryabchikova}, {Stempels}
   \& {Weiss}}{{Kupka} et~al.}{1999}]{Kup99}
{Kupka} F.,  {Piskunov} N.,  {Ryabchikova} T.~A.,  {Stempels} H.~C.,   {Weiss}
  W.~W.,  1999, \mn@doi [\aaps] {10.1051/aas:1999267}, \href
  {http://adsabs.harvard.edu/abs/1999A%26AS..138..119K} {138, 119}

\bibitem[\protect\citeauthoryear{{Kupka}, {Ryabchikova}, {Piskunov}, {Stempels}
   \& {Weiss}}{{Kupka} et~al.}{2000}]{Kup00}
{Kupka} F.~G.,  {Ryabchikova} T.~A.,  {Piskunov} N.~E.,  {Stempels} H.~C.,
  {Weiss} W.~W.,  2000, Baltic Astronomy, \href
  {http://adsabs.harvard.edu/abs/2000BaltA...9..590K} {9, 590}

\bibitem[\protect\citeauthoryear{{L{\`e}bre}, {de Laverny}, {Do Nascimento}  \&
  {de Medeiros}}{{L{\`e}bre} et~al.}{2006}]{Leb06}
{L{\`e}bre} A.,  {de Laverny} P.,  {Do Nascimento} J.~D. J.,   {de Medeiros}
  J.~R.,  2006, \mn@doi [\aap] {10.1051/0004-6361:20053485}, \href
  {https://ui.adsabs.harvard.edu/abs/2006A&A...450.1173L} {450, 1173}

\bibitem[\protect\citeauthoryear{{Lebzelter} et~al.,}{{Lebzelter}
  et~al.}{2012}]{Leb12}
{Lebzelter} T.,  et~al., 2012, \mn@doi [\aap] {10.1051/0004-6361/201117728},
  \href {https://ui.adsabs.harvard.edu/abs/2012A&A...539A.109L} {539, A109}

\bibitem[\protect\citeauthoryear{{Liu}, {Zhao}, {Shi}, {Pietrzy{\'n}ski}  \&
  {Gieren}}{{Liu} et~al.}{2007}]{Liu07}
{Liu} Y.~J.,  {Zhao} G.,  {Shi} J.~R.,  {Pietrzy{\'n}ski} G.,   {Gieren} W.,
  2007, \mn@doi [\mnras] {10.1111/j.1365-2966.2007.11852.x}, \href
  {http://adsabs.harvard.edu/abs/2007MNRAS.382..553L} {382, 553}

\bibitem[\protect\citeauthoryear{{Luck}}{{Luck}}{2014}]{Luc14}
{Luck} R.~E.,  2014, \mn@doi [\aj] {10.1088/0004-6256/147/6/137}, \href
  {http://adsabs.harvard.edu/abs/2014AJ....147..137L} {147, 137}

\bibitem[\protect\citeauthoryear{{Lyubimkov}, {Lambert}, {Rostopchin},
  {Rachkovskaya}  \& {Poklad}}{{Lyubimkov} et~al.}{2010}]{Lyu10}
{Lyubimkov} L.~S.,  {Lambert} D.~L.,  {Rostopchin} S.~I.,  {Rachkovskaya}
  T.~M.,   {Poklad} D.~B.,  2010, \mn@doi [\mnras]
  {10.1111/j.1365-2966.2009.15979.x}, \href
  {http://adsabs.harvard.edu/abs/2010MNRAS.402.1369L} {402, 1369}

\bibitem[\protect\citeauthoryear{{Lyubimkov}, {Lambert}, {Kaminsky},
  {Pavlenko}, {Poklad}  \& {Rachkovskaya}}{{Lyubimkov} et~al.}{2012}]{Lyu12}
{Lyubimkov} L.~S.,  {Lambert} D.~L.,  {Kaminsky} B.~M.,  {Pavlenko} Y.~V.,
  {Poklad} D.~B.,   {Rachkovskaya} T.~M.,  2012, \mn@doi [\mnras]
  {10.1111/j.1365-2966.2012.21617.x}, \href
  {https://ui.adsabs.harvard.edu/abs/2012MNRAS.427...11L} {427, 11}

\bibitem[\protect\citeauthoryear{{Majewski} et~al.,}{{Majewski}
  et~al.}{2017}]{Maj17}
{Majewski} S.~R.,  et~al., 2017, \mn@doi [\aj] {10.3847/1538-3881/aa784d},
  \href {https://ui.adsabs.harvard.edu/abs/2017AJ....154...94M} {154, 94}

\bibitem[\protect\citeauthoryear{{Maldonado}, {Mart{\'{\i}}nez-Arn{\'a}iz},
  {Eiroa}, {Montes}  \& {Montesinos}}{{Maldonado} et~al.}{2010}]{Mal10}
{Maldonado} J.,  {Mart{\'{\i}}nez-Arn{\'a}iz} R.~M.,  {Eiroa} C.,  {Montes} D.,
    {Montesinos} B.,  2010, \mn@doi [\aap] {10.1051/0004-6361/201014948}, \href
  {http://adsabs.harvard.edu/abs/2010A%26A...521A..12M} {521, A12}

\bibitem[\protect\citeauthoryear{{Mann} et~al.,}{{Mann} et~al.}{2019}]{Man19}
{Mann} A.~W.,  et~al., 2019, \mn@doi [\apj] {10.3847/1538-4357/aaf3bc}, \href
  {https://ui.adsabs.harvard.edu/abs/2019ApJ...871...63M} {871, 63}

\bibitem[\protect\citeauthoryear{{Mart{\'{\i}}nez-Arn{\'a}iz}, {Maldonado},
  {Montes}, {Eiroa}  \& {Montesinos}}{{Mart{\'{\i}}nez-Arn{\'a}iz}
  et~al.}{2010}]{Mar10}
{Mart{\'{\i}}nez-Arn{\'a}iz} R.,  {Maldonado} J.,  {Montes} D.,  {Eiroa} C.,
  {Montesinos} B.,  2010, \mn@doi [\aap] {10.1051/0004-6361/200913725}, \href
  {http://adsabs.harvard.edu/abs/2010A%26A...520A..79M} {520, A79}

\bibitem[\protect\citeauthoryear{{Massarotti}, {Latham}, {Stefanik}  \&
  {Fogel}}{{Massarotti} et~al.}{2008}]{Mas08}
{Massarotti} A.,  {Latham} D.~W.,  {Stefanik} R.~P.,   {Fogel} J.,  2008,
  \mn@doi [\aj] {10.1088/0004-6256/135/1/209}, \href
  {http://adsabs.harvard.edu/abs/2008AJ....135..209M} {135, 209}

\bibitem[\protect\citeauthoryear{{Mayor} et~al.,}{{Mayor} et~al.}{2003}]{May03}
{Mayor} M.,  et~al., 2003, The Messenger, \href
  {https://ui.adsabs.harvard.edu/abs/2003Msngr.114...20M} {114, 20}

\bibitem[\protect\citeauthoryear{{McCarthy} \& {Wilhelm}}{{McCarthy} \&
  {Wilhelm}}{2014}]{McC14}
{McCarthy} K.,  {Wilhelm} R.~J.,  2014, \mn@doi [\aj]
  {10.1088/0004-6256/148/4/70}, \href
  {http://cdsads.u-strasbg.fr/abs/2014AJ....148...70M} {148, 70}

\bibitem[\protect\citeauthoryear{{McWilliam}}{{McWilliam}}{1990}]{McW90}
{McWilliam} A.,  1990, \mn@doi [\apjs] {10.1086/191527}, \href
  {http://adsabs.harvard.edu/abs/1990ApJS...74.1075M} {74, 1075}

\bibitem[\protect\citeauthoryear{{Mel{\'e}ndez} \& {Barbuy}}{{Mel{\'e}ndez} \&
  {Barbuy}}{2009}]{Mel09}
{Mel{\'e}ndez} J.,  {Barbuy} B.,  2009, \mn@doi [\aap]
  {10.1051/0004-6361/200811508}, \href
  {http://adsabs.harvard.edu/abs/2009A%26A...497..611M} {497, 611}

\bibitem[\protect\citeauthoryear{{Montes} et~al.,}{{Montes}
  et~al.}{2018}]{Mon18}
{Montes} D.,  et~al., 2018, \mn@doi [\mnras] {10.1093/mnras/sty1295}, \href
  {http://adsabs.harvard.edu/abs/2018MNRAS.479.1332M} {479, 1332}

\bibitem[\protect\citeauthoryear{{Morel} et~al.,}{{Morel} et~al.}{2014}]{Mor14}
{Morel} T.,  et~al., 2014, \mn@doi [\aap] {10.1051/0004-6361/201322810}, \href
  {http://adsabs.harvard.edu/abs/2014A%26A...564A.119M} {564, A119}

\bibitem[\protect\citeauthoryear{{Morton}}{{Morton}}{2000}]{Mor00}
{Morton} D.~C.,  2000, \mn@doi [\apjs] {10.1086/317349}, \href
  {http://adsabs.harvard.edu/abs/2000ApJS..130..403M} {130, 403}

\bibitem[\protect\citeauthoryear{{Mucciarelli}, {Pancino}, {Lovisi}, {Ferraro}
  \& {Lapenna}}{{Mucciarelli} et~al.}{2013}]{Muc13}
{Mucciarelli} A.,  {Pancino} E.,  {Lovisi} L.,  {Ferraro} F.~R.,   {Lapenna}
  E.,  2013, \mn@doi [\apj] {10.1088/0004-637X/766/2/78}, \href
  {http://adsabs.harvard.edu/abs/2013ApJ...766...78M} {766, 78}

\bibitem[\protect\citeauthoryear{{Nagel} et~al.,}{{Nagel} et~al.}{2019}]{Nag20}
{Nagel} E.,  et~al., 2019, \aap, submitted

\bibitem[\protect\citeauthoryear{{Nicholls} et~al.,}{{Nicholls}
  et~al.}{2017}]{Nic17}
{Nicholls} C.~P.,  et~al., 2017, \mn@doi [\aap] {10.1051/0004-6361/201629244},
  \href {https://ui.adsabs.harvard.edu/abs/2017A&A...598A..79N} {598, A79}

\bibitem[\protect\citeauthoryear{{Nidever}, {Marcy}, {Butler}, {Fischer}  \&
  {Vogt}}{{Nidever} et~al.}{2002}]{Nid02}
{Nidever} D.~L.,  {Marcy} G.~W.,  {Butler} R.~P.,  {Fischer} D.~A.,   {Vogt}
  S.~S.,  2002, \mn@doi [\apjs] {10.1086/340570}, \href
  {http://adsabs.harvard.edu/abs/2002ApJS..141..503N} {141, 503}

\bibitem[\protect\citeauthoryear{{Oliva}, {Sanna}, {Rainer}, {Massi}, {Tozzi}
  \& {Origlia}}{{Oliva} et~al.}{2018}]{Oli18}
{Oliva} E.,  {Sanna} N.,  {Rainer} M.,  {Massi} F.,  {Tozzi} A.,   {Origlia}
  L.,  2018, in \procspie. p. 1070274, \mn@doi{10.1117/12.2309927}

\bibitem[\protect\citeauthoryear{{Origlia} et~al.,}{{Origlia}
  et~al.}{2014}]{Ori14}
{Origlia} L.,  et~al., 2014, in \procspie. p. 91471E,
  \mn@doi{10.1117/12.2054743}

\bibitem[\protect\citeauthoryear{{Park} et~al.,}{{Park} et~al.}{2018}]{Par18}
{Park} S.,  et~al., 2018, \mn@doi [\apjs] {10.3847/1538-4365/aadd14}, \href
  {https://ui.adsabs.harvard.edu/abs/2018ApJS..238...29P} {238, 29}

\bibitem[\protect\citeauthoryear{{Passegger} et~al.,}{{Passegger}
  et~al.}{2019}]{Pas19}
{Passegger} V.~M.,  et~al., 2019, \mn@doi [\aap] {10.1051/0004-6361/201935679},
  \href {https://ui.adsabs.harvard.edu/abs/2019A&A...627A.161P} {627, A161}

\bibitem[\protect\citeauthoryear{{Pavlenko}, {Jenkins}, {Jones}, {Ivanyuk}  \&
  {Pinfield}}{{Pavlenko} et~al.}{2012}]{Pav12}
{Pavlenko} Y.~V.,  {Jenkins} J.~S.,  {Jones} H.~R.~A.,  {Ivanyuk} O.,
  {Pinfield} D.~J.,  2012, \mn@doi [\mnras] {10.1111/j.1365-2966.2012.20629.x},
  \href {https://ui.adsabs.harvard.edu/abs/2012MNRAS.422..542P} {422, 542}

\bibitem[\protect\citeauthoryear{{Piskunov} \& {Valenti}}{{Piskunov} \&
  {Valenti}}{2002}]{Pis02}
{Piskunov} N.~E.,  {Valenti} J.~A.,  2002, \mn@doi [\aap]
  {10.1051/0004-6361:20020175}, \href
  {https://ui.adsabs.harvard.edu/abs/2002A&A...385.1095P} {385, 1095}

\bibitem[\protect\citeauthoryear{{Piskunov} \& {Valenti}}{{Piskunov} \&
  {Valenti}}{2017}]{Pis17}
{Piskunov} N.,  {Valenti} J.~A.,  2017, \mn@doi [\aap]
  {10.1051/0004-6361/201629124}, \href
  {https://ui.adsabs.harvard.edu/abs/2017A&A...597A..16P} {597, A16}

\bibitem[\protect\citeauthoryear{{Piskunov}, {Kupka}, {Ryabchikova}, {Weiss}
  \& {Jeffery}}{{Piskunov} et~al.}{1995}]{Pis95}
{Piskunov} N.~E.,  {Kupka} F.,  {Ryabchikova} T.~A.,  {Weiss} W.~W.,
  {Jeffery} C.~S.,  1995, \aaps, \href
  {http://adsabs.harvard.edu/abs/1995A%26AS..112..525P} {112, 525}

\bibitem[\protect\citeauthoryear{{Pourbaix} et~al.,}{{Pourbaix}
  et~al.}{2004}]{Pou04}
{Pourbaix} D.,  et~al., 2004, \mn@doi [\aap] {10.1051/0004-6361:20041213},
  \href {http://adsabs.harvard.edu/abs/2004A%26A...424..727P} {424, 727}

\bibitem[\protect\citeauthoryear{{Press}, {Teukolsky}, {Vetterling}  \&
  {Flannery}}{{Press} et~al.}{2002}]{Pre02}
{Press} W.~H.,  {Teukolsky} S.~A.,  {Vetterling} W.~T.,   {Flannery} B.~P.,
  2002, {Numerical recipes in C++ : the art of scientific computing}.
Cambridge University Press

\bibitem[\protect\citeauthoryear{{Quirrenbach} et~al.,}{{Quirrenbach}
  et~al.}{2018}]{Qui18}
{Quirrenbach} A.,  et~al., 2018, in Ground-based and Airborne Instrumentation
  for Astronomy VII. p. 107020W, \mn@doi{10.1117/12.2313689}

\bibitem[\protect\citeauthoryear{{Reiners} et~al.,}{{Reiners}
  et~al.}{2018}]{Rei18}
{Reiners} A.,  et~al., 2018, \mn@doi [\aap] {10.1051/0004-6361/201732054},
  \href {https://ui.adsabs.harvard.edu/abs/2018A&A...612A..49R} {612, A49}

\bibitem[\protect\citeauthoryear{{Rodrigues} et~al.,}{{Rodrigues}
  et~al.}{2014}]{Rod14}
{Rodrigues} T.~S.,  et~al., 2014, \mn@doi [\mnras] {10.1093/mnras/stu1907},
  \href {https://ui.adsabs.harvard.edu/abs/2014MNRAS.445.2758R} {445, 2758}

\bibitem[\protect\citeauthoryear{{Rodrigues} et~al.,}{{Rodrigues}
  et~al.}{2017}]{Rod17}
{Rodrigues} T.~S.,  et~al., 2017, \mn@doi [\mnras] {10.1093/mnras/stx120},
  \href {https://ui.adsabs.harvard.edu/abs/2017MNRAS.467.1433R} {467, 1433}

\bibitem[\protect\citeauthoryear{{Ryabchikova}, {Piskunov}, {Kurucz},
  {Stempels}, {Heiter}, {Pakhomov}  \& {Barklem}}{{Ryabchikova}
  et~al.}{2015}]{Rya15}
{Ryabchikova} T.,  {Piskunov} N.,  {Kurucz} R.~L.,  {Stempels} H.~C.,  {Heiter}
  U.,  {Pakhomov} Y.,   {Barklem} P.~S.,  2015, \mn@doi [\physscr]
  {10.1088/0031-8949/90/5/054005}, \href
  {http://adsabs.harvard.edu/abs/2015PhyS...90e4005R} {90, 054005}

\bibitem[\protect\citeauthoryear{{Santos} et~al.,}{{Santos}
  et~al.}{2013}]{San13}
{Santos} N.~C.,  et~al., 2013, \mn@doi [\aap] {10.1051/0004-6361/201321286},
  \href {https://ui.adsabs.harvard.edu/abs/2013A&A...556A.150S} {556, A150}

\bibitem[\protect\citeauthoryear{{Scarfe}, {Funakawa}, {Delaney}  \&
  {Barlow}}{{Scarfe} et~al.}{1983}]{Sca83}
{Scarfe} C.~D.,  {Funakawa} H.,  {Delaney} P.~A.,   {Barlow} D.~J.,  1983,
  \jrasc, \href {https://ui.adsabs.harvard.edu/abs/1983JRASC..77..126S} {77,
  126}

\bibitem[\protect\citeauthoryear{{Schweitzer} et~al.,}{{Schweitzer}
  et~al.}{2019}]{Sch19}
{Schweitzer} A.,  et~al., 2019, \mn@doi [\aap] {10.1051/0004-6361/201834965},
  \href {https://ui.adsabs.harvard.edu/abs/2019A&A...625A..68S} {625, A68}

\bibitem[\protect\citeauthoryear{{Skrutskie} et~al.,}{{Skrutskie}
  et~al.}{2006}]{Skr06}
{Skrutskie} M.~F.,  et~al., 2006, \mn@doi [\aj] {10.1086/498708}, \href
  {http://adsabs.harvard.edu/abs/2006AJ....131.1163S} {131, 1163}

\bibitem[\protect\citeauthoryear{{Smette} et~al.,}{{Smette}
  et~al.}{2015}]{Sme15}
{Smette} A.,  et~al., 2015, \mn@doi [\aap] {10.1051/0004-6361/201423932}, \href
  {https://ui.adsabs.harvard.edu/abs/2015A&A...576A..77S} {576, A77}

\bibitem[\protect\citeauthoryear{{Sneden}}{{Sneden}}{1973}]{Sne73}
{Sneden} C.~A.,  1973, PhD thesis, The University of Texas at Austin

\bibitem[\protect\citeauthoryear{{Soubiran}, {Jasniewicz}, {Chemin}, {Crifo},
  {Udry}, {Hestroffer}  \& {Katz}}{{Soubiran} et~al.}{2013}]{Sou13}
{Soubiran} C.,  {Jasniewicz} G.,  {Chemin} L.,  {Crifo} F.,  {Udry} S.,
  {Hestroffer} D.,   {Katz} D.,  2013, \mn@doi [\aap]
  {10.1051/0004-6361/201220927}, \href
  {http://adsabs.harvard.edu/abs/2013A%26A...552A..64S} {552, A64}

\bibitem[\protect\citeauthoryear{{Soubiran}, {Le Campion}, {Brouillet}  \&
  {Chemin}}{{Soubiran} et~al.}{2016}]{Sou16}
{Soubiran} C.,  {Le Campion} J.-F.,  {Brouillet} N.,   {Chemin} L.,  2016,
  \mn@doi [\aap] {10.1051/0004-6361/201628497}, \href
  {http://adsabs.harvard.edu/abs/2016A%26A...591A.118S} {591, A118}

\bibitem[\protect\citeauthoryear{{Sousa}, {Santos}, {Israelian}, {Mayor}  \&
  {Monteiro}}{{Sousa} et~al.}{2007}]{Sou07}
{Sousa} S.~G.,  {Santos} N.~C.,  {Israelian} G.,  {Mayor} M.,   {Monteiro}
  M.~J.~P.~F.~G.,  2007, \mn@doi [\aap] {10.1051/0004-6361:20077288}, \href
  {http://adsabs.harvard.edu/abs/2007A%26A...469..783S} {469, 783}

\bibitem[\protect\citeauthoryear{{Sousa} et~al.,}{{Sousa} et~al.}{2008}]{Sou08}
{Sousa} S.~G.,  et~al., 2008, \mn@doi [\aap] {10.1051/0004-6361:200809698},
  \href {http://adsabs.harvard.edu/abs/2008A%26A...487..373S} {487, 373}

\bibitem[\protect\citeauthoryear{{Spada}, {Demarque}, {Kim}, {Boyajian}  \&
  {Brewer}}{{Spada} et~al.}{2017}]{Spa17}
{Spada} F.,  {Demarque} P.,  {Kim} Y.-C.,  {Boyajian} T.~S.,   {Brewer} J.~M.,
  2017, \mn@doi [\apj] {10.3847/1538-4357/aa661d}, \href
  {http://adsabs.harvard.edu/abs/2017ApJ...838..161S} {838, 161}

\bibitem[\protect\citeauthoryear{Spencer~Jones \& Furner}{Spencer~Jones \&
  Furner}{1937}]{Spe37}
Spencer~Jones H.,  Furner H.~H.,  1937, \mn@doi [Monthly Notices of the Royal
  Astronomical Society] {10.1093/mnras/98.2.92}, 98, 92

\bibitem[\protect\citeauthoryear{{Steinmetz} et~al.,}{{Steinmetz}
  et~al.}{2006}]{Ste06}
{Steinmetz} M.,  et~al., 2006, \mn@doi [\aj] {10.1086/506564}, \href
  {http://adsabs.harvard.edu/abs/2006AJ....132.1645S} {132, 1645}

\bibitem[\protect\citeauthoryear{{Tabernero}, {Montes}  \& {Gonz{\'a}lez
  Hern{\'a}ndez}}{{Tabernero} et~al.}{2012}]{Tab12}
{Tabernero} H.~M.,  {Montes} D.,   {Gonz{\'a}lez Hern{\'a}ndez} J.~I.,  2012,
  \mn@doi [\aap] {10.1051/0004-6361/201117506}, \href
  {http://adsabs.harvard.edu/abs/2012A%26A...547A..13T} {547, A13}

\bibitem[\protect\citeauthoryear{{Tabernero}, {Montes}, {Gonz{\'a}lez
  Hern{\'a}ndez}  \& {Ammler-von Eiff}}{{Tabernero} et~al.}{2017}]{Tab17}
{Tabernero} H.~M.,  {Montes} D.,  {Gonz{\'a}lez Hern{\'a}ndez} J.~I.,
  {Ammler-von Eiff} M.,  2017, \mn@doi [\aap] {10.1051/0004-6361/201322526},
  \href {https://ui.adsabs.harvard.edu/abs/2017A&A...597A..33T} {597, A33}

\bibitem[\protect\citeauthoryear{{Tabernero}, {Dorda}, {Negueruela}  \&
  {Gonz{\'a}lez-Fern{\'a}ndez}}{{Tabernero} et~al.}{2018}]{Tab18}
{Tabernero} H.~M.,  {Dorda} R.,  {Negueruela} I.,
  {Gonz{\'a}lez-Fern{\'a}ndez} C.,  2018, \mn@doi [\mnras]
  {10.1093/mnras/sty399}, \href
  {http://adsabs.harvard.edu/abs/2018MNRAS.476.3106T} {476, 3106}

\bibitem[\protect\citeauthoryear{{Tabernero}, {Marfil}, {Montes}  \&
  {Gonz{\'a}lez Hern{\'a}ndez}}{{Tabernero} et~al.}{2019}]{Tab19}
{Tabernero} H.~M.,  {Marfil} E.,  {Montes} D.,   {Gonz{\'a}lez Hern{\'a}ndez}
  J.~I.,  2019, \aap, \href
  {https://ui.adsabs.harvard.edu/abs/2019arXiv190706512T} {628, A131}

\bibitem[\protect\citeauthoryear{{Takeda} et~al.,}{{Takeda}
  et~al.}{2005}]{Tak05}
{Takeda} Y.,  et~al., 2005, \mn@doi [\pasj] {10.1093/pasj/57.1.13}, \href
  {http://adsabs.harvard.edu/abs/2005PASJ...57...13T} {57, 13}

\bibitem[\protect\citeauthoryear{{Takeda}, {Sato}  \& {Murata}}{{Takeda}
  et~al.}{2008}]{Tak08}
{Takeda} Y.,  {Sato} B.,   {Murata} D.,  2008, \mn@doi [\pasj]
  {10.1093/pasj/60.4.781}, \href
  {https://ui.adsabs.harvard.edu/abs/2008PASJ...60..781T} {60, 781}

\bibitem[\protect\citeauthoryear{{Thygesen} et~al.,}{{Thygesen}
  et~al.}{2012}]{Thy12}
{Thygesen} A.~O.,  et~al., 2012, \mn@doi [\aap] {10.1051/0004-6361/201219237},
  \href {http://adsabs.harvard.edu/abs/2012A%26A...543A.160T} {543, A160}

\bibitem[\protect\citeauthoryear{{Tsantaki}, {Sousa}, {Adibekyan}, {Santos},
  {Mortier}  \& {Israelian}}{{Tsantaki} et~al.}{2013}]{Tsa13}
{Tsantaki} M.,  {Sousa} S.~G.,  {Adibekyan} V.~Z.,  {Santos} N.~C.,  {Mortier}
  A.,   {Israelian} G.,  2013, \mn@doi [\aap] {10.1051/0004-6361/201321103},
  \href {http://adsabs.harvard.edu/abs/2013A%26A...555A.150T} {555, A150}

\bibitem[\protect\citeauthoryear{{Valenti} \& {Fischer}}{{Valenti} \&
  {Fischer}}{2005}]{Val05}
{Valenti} J.~A.,  {Fischer} D.~A.,  2005, \mn@doi [\apjs] {10.1086/430500},
  \href {http://adsabs.harvard.edu/abs/2005ApJS..159..141V} {159, 141}

\bibitem[\protect\citeauthoryear{{Wildi} et~al.,}{{Wildi} et~al.}{2017}]{Wil17}
{Wildi} F.,  et~al., 2017, in \procspie. p. 1040018,
  \mn@doi{10.1117/12.2275660}

\bibitem[\protect\citeauthoryear{{Worek} \& {Beardsley}}{{Worek} \&
  {Beardsley}}{1977}]{Wor77}
{Worek} T.~F.,  {Beardsley} W.~R.,  1977, \mn@doi [\apj] {10.1086/155562},
  \href {https://ui.adsabs.harvard.edu/abs/1977ApJ...217..134W} {217, 134}

\bibitem[\protect\citeauthoryear{{Wright} et~al.,}{{Wright}
  et~al.}{2018}]{Wri18}
{Wright} J.~T.,  et~al., 2018, in American Astronomical Society Meeting
  Abstracts \#231. p. 246.45

\bibitem[\protect\citeauthoryear{{Wu}, {Singh}, {Prugniel}, {Gupta}  \&
  {Koleva}}{{Wu} et~al.}{2011}]{Wu11}
{Wu} Y.,  {Singh} H.~P.,  {Prugniel} P.,  {Gupta} R.,   {Koleva} M.,  2011,
  \mn@doi [\aap] {10.1051/0004-6361/201015014}, \href
  {http://cdsads.u-strasbg.fr/abs/2011A%26A...525A..71W} {525, A71}

\bibitem[\protect\citeauthoryear{{Zamora} et~al.,}{{Zamora}
  et~al.}{2015}]{Zam15}
{Zamora} O.,  et~al., 2015, \mn@doi [\aj] {10.1088/0004-6256/149/6/181}, \href
  {https://ui.adsabs.harvard.edu/abs/2015AJ....149..181Z} {149, 181}

\bibitem[\protect\citeauthoryear{{Zechmeister}, {Anglada-Escud{\'e}}  \&
  {Reiners}}{{Zechmeister} et~al.}{2014}]{Zec14}
{Zechmeister} M.,  {Anglada-Escud{\'e}} G.,   {Reiners} A.,  2014, \mn@doi
  [\aap] {10.1051/0004-6361/201322746}, \href
  {https://ui.adsabs.harvard.edu/abs/2014A&A...561A..59Z} {561, A59}

\bibitem[\protect\citeauthoryear{{da Silva} et~al.,}{{da Silva}
  et~al.}{2006}]{Sil06}
{da Silva} L.,  et~al., 2006, \mn@doi [\aap] {10.1051/0004-6361:20065105},
  \href {https://ui.adsabs.harvard.edu/abs/2006A&A...458..609D} {458, 609}

\bibitem[\protect\citeauthoryear{{da Silva}, {Milone}  \& {Rocha-Pinto}}{{da
  Silva} et~al.}{2015}]{Sil15}
{da Silva} R.,  {Milone} A.~d.~C.,   {Rocha-Pinto} H.~J.,  2015, \mn@doi [\aap]
  {10.1051/0004-6361/201525770}, \href
  {http://adsabs.harvard.edu/abs/2015A%26A...580A..24D} {580, A24}

\bibitem[\protect\citeauthoryear{{dos Santos} et~al.,}{{dos Santos}
  et~al.}{2016}]{dos16}
{dos Santos} L.~A.,  et~al., 2016, \mn@doi [\aap]
  {10.1051/0004-6361/201628558}, \href
  {http://adsabs.harvard.edu/abs/2016A%26A...592A.156D} {592, A156}

\bibitem[\protect\citeauthoryear{{van Leeuwen}}{{van Leeuwen}}{2007}]{Lee07}
{van Leeuwen} F.,  2007, \mn@doi [\aap] {10.1051/0004-6361:20078357}, \href
  {https://ui.adsabs.harvard.edu/abs/2007A&A...474..653V} {474, 653}

\makeatother
\end{thebibliography}




\appendix

\section{Appendix}
In Table~\ref{tab:par_stars_ref} we give the literature values of the stellar atmospheric parameters for the selected sample. In Tables~\ref{tab:par_stars_stepar} and \ref{tab:par_stars_stepar_vis} we give the stellar atmospheric parameters computed with {\sc StePar} in the whole VIS+NIR region and VIS region, respectively. In Tables~\ref{tab:line_table_all_fe_i} and \ref{tab:line_table_all_fe_ii} we list the \ion{Fe}{i} and \ion{Fe}{ii} lines along with their parameters, respectively, for metal-rich dwarfs (MRD), metal-poor dwarfs (MPD), metal-rich giants (MRG), and metal-poor giants (MPG). Finally, we include the CARMENES spectrum of the reference, metal-rich dwarf 18~Sco in Fig.~\ref{fig:18Scospectrum}, along with the \ion{Fe}{i} and \ion{Fe}{ii} lines indicated in red and green, respectively. 

\clearpage
\onecolumn
{
\footnotesize
\begin{landscape}
\begin{longtable}{p{1.3cm}p{0.7cm}cccllclcp{0.6cm}ccccl}

\caption{\label{tab:par_stars_ref} Reference stellar parameters of the selected CARMENES sample.}\\
\hline\noalign{\smallskip}
Name & HD & $\alpha$ (J2000) & $\delta$ (J2000) & $\pi$ & Ref.$^a$ & SpT$^b$ & $\varv_{\rm r}$    & Ref.$^c$ & $\varv\sin{i}$     & Ref.$^d$ & $T_{\rm eff}$ & $\log{g}$ & $\xi$          & [Fe/H] & Ref.$^e$\\
     &    &                  &                  & [mas] &          &         & [km\,s$^{-1}$]     &          & [km\,s$^{-1}$]     &          & [K]           & [dex]     & [km\,s$^{-1}$] & [dex]  &         \\
\noalign{\smallskip}\hline\noalign{\smallskip}
\endfirsthead

\caption{Reference stellar parameters of the selected CARMENES sample (cont.).}\\
\hline\noalign{\smallskip}
Name & HD & $\alpha$ (J2000) & $\delta$ (J2000) & $\pi$ & Ref.$^a$ & SpT$^b$ & $\varv_{\rm r}$    & Ref.$^c$ & $\varv\sin{i}$     & Ref.$^d$ & $T_{\rm eff}$ & $\log{g}$ & $\xi$          & [Fe/H] & Ref.$^e$\\
     &    &                  &                  & [mas] &          &         & [km\,s$^{-1}$]     &          & [km\,s$^{-1}$]     &          & [K]           & [dex]     & [km\,s$^{-1}$] & [dex]  &         \\
\noalign{\smallskip}\hline\noalign{\smallskip}
\endhead

\noalign{\smallskip}\hline
\endfoot

\noalign{\smallskip}\hline\noalign{\smallskip}
\multicolumn{16}{p{21.8cm}}{\footnotesize $^a$\,References for parallax, $\pi$: G:{\it Gaia} DR2 \citep{gaia}; H: Hipparcos \citep{Lee07}.}\\
\multicolumn{16}{p{21.8cm}}{\footnotesize $^b$\,Reference for spectral type, SpT: Caballero et al. (in prep.).}\\
\multicolumn{16}{p{21.8cm}}{\footnotesize $^c$\,References for $\varv_{\rm r}$: 
                          Fam05: \citet{Fam05}; Gaia:  \citet{gaia};  Gon06: \citet{Gon06}; Hay18: \citet{Hay18};
                          Kar04: \citet{Kar04}; Jof15: \citet{Jof15}; Mal10: \citet{Mal10}; Mas08: \citet{Mas08}; 
                          Mon18: \citet{Mon18}; Nid02: \citet{Nid02}; Pou04: \citet{Pou04}; Sou13: \citet{Sou13}.}\\
\multicolumn{16}{p{21.8cm}}{\footnotesize $^d$\,References for $\varv\sin{i}$:
                          San16: \citet{dos16}; Jof15: \citet{Jof15}; Gra86: \citet{Gra86}; Gra18: \citet{Gra18}; 
                          Leb06: \citet{Leb06}; Lyu12: \citet{Lyu12}; Mar10: \citet{Mar10}; Mas08: \citet{Mas08}; 
                          McC14: \citet{McC14};  Med00: \citet{Med00}; Med02: \citet{Med02}; Pav12: \citet{Pav12};
                          Tak05: \citet{Tak05}.}\\
\multicolumn{16}{p{21.8cm}}{\footnotesize $^e$\,References for $T_{\rm eff}$, $\log{g}$, $\xi$, and [Fe/H]:
                          All04: \citet{All04}; Hei03: \citet{Hei03}; Hek07: \citet{Hek07}; Jof14: \citet{Jof14}; 
                          Jof15: \citet{Jof15}; Jof18: \citet{Jof18}; Liu07: \citet{Liu07}; Luc14: \citet{Luc14}; 
                          Lyu10: \citet{Lyu10}; McW90: \citet{McW90}; Mor14: \citet{Mor14}; San13: \citet{San13}; 
                          Sil15: \citet{Sil15}; Sou08: \citet{Sou08}; Tak08: \citet{Tak08};  Thy12: \citet{Thy12};
                          Val05: \citet{Val05}; Wu11: \citet{Wu11}.}\\
\multicolumn{16}{p{21.8cm}}{$^{f}$ \emph{Gaia} benchmark star \citep{Jof14, Jof18, Hei15}.}\\
\endlastfoot

\multicolumn{16}{c}{Metal-rich dwarfs (MRD)}\\
\hline\noalign{\smallskip}
Sun$^{f}$            & ...    & ...         & ...           & ...      $\pm$ ...    & ... & G2\,V     &     0.00 & ...   & 1.6  & Pav12 & 5771 $\pm$ 1   & 4.44 $\pm$ 0.00 & 1.20 $\pm$ 0.18 & $+$0.03 $\pm$ 0.05 & Jof18 \\
HD~3765              & 3765   & 00 40 49.29 & $+$40 11 13.3 & 55.7562  $\pm$ 0.1002 & G & K2\,V       & $-$63.11 & Sou13 & 9.2  & Mar10 & 5032 $\pm$ 44  & 4.59 $\pm$ 0.06 & 0.85 $\pm$ ...  & $+$0.18 $\pm$ 0.03 & Val05 \\
HD~100167            & 100167 & 11 31 53.92 & $+$41 26 21.5 & 28.4793  $\pm$ 0.1513 & G & F8\,V       & $-$29.49 & Nid02 & 5.0  & McC14 & 5915 $\pm$ 44  & 4.38 $\pm$ 0.06 & 0.85 $\pm$ ...  & $+$0.06 $\pm$ 0.03 & Val05 \\
61~UMa               & 101501 & 11 41 03.01 & $+$34 12 06.4 & 104.3904 $\pm$ 0.1287 & G & G8\,V       &  $-$5.18 & Nid02 & 3.3  & Mar10 & 5488 $\pm$ 44  & 4.43 $\pm$ 0.06 & 0.85 $\pm$ ...  & $-$0.03 $\pm$ 0.03 & Val05 \\
$\beta$~Vir$^{f}$    & 102870 & 11 50 41.73 & $+$01 45 52.8 & 89.9258  $\pm$ 0.5195 & G & F9\,V       &  $+$4.71 & Nid02 & 3.4  & Mar10 & 6083 $\pm$ 41  & 4.10 $\pm$ 0.02 & 1.40 $\pm$ 0.09 & $+$0.24 $\pm$ 0.07 & Jof18 \\
$\beta$~CVn          & 109358 & 12 33 44.54 & $+$41 21 27.0 & 116.1298 $\pm$ 0.6776 & G & G0\,V       &  $+$6.52 & Nid02 & 3.2  & Mar10 & 5930 $\pm$ 44  & 4.44 $\pm$ 0.06 & 0.85 $\pm$ ...  & $-$0.16 $\pm$ 0.03 & Val05 \\
$\beta$~Com          & 114710 & 13 11 52.38 & $+$27 52 41.1 & 108.8951 $\pm$ 0.3487 & G & F9.5\,V     &  $+$5.46 & Nid02 & 4.7  & Mar10 & 6075 $\pm$ 44  & 4.57 $\pm$ 0.06 & 0.85 $\pm$ ...  & $+$0.07 $\pm$ 0.03 & Val05 \\
\end{longtable}
\end{landscape}
}
\clearpage
\twocolumn

\clearpage
\onecolumn
\begin{longtable}{lccccccc}

\caption{\label{tab:par_stars_stepar} Stellar atmospheric parameters of the selected sample under {\sc StePar} and $\log{g}$ values obtained
with PARAM assuming parallaxes from {\emph Gaia} DR2 and the Hipparcos mission.}\\
\hline\noalign{\smallskip}
Name & $\varv_{\rm r}$    & S/N & $T_{\rm eff}$ & $\log{g}$ & $\log{g}_{\rm PARAM}$ & $\xi$         & [Fe/H] \\
     & [km\,s$^{-1}$]     &     & [K]           & [dex]     & [dex]                 & [km\,s$^{-1}$] & [dex]  \\
\noalign{\smallskip}\hline\noalign{\smallskip}
\endfirsthead

\caption{Stellar atmospheric parameters of the selected sample under {\sc StePar} and $\log{g}$ values obtained
with PARAM assuming parallaxes from {\emph Gaia} DR2 and the Hipparcos mission (cont.).}\\
\hline\noalign{\smallskip}
Name & $\varv_{\rm r}$    & S/N & $T_{\rm eff}$ & $\log{g}$ & $\log{g}_{\rm PARAM}$ & $\xi$         & [Fe/H]\\
     & [km\,s$^{-1}$]     &     & [K]           & [dex]     & [dex]                 & [km\,s$^{-1}$] & [dex]  \\
\noalign{\smallskip}\hline\noalign{\smallskip}
\endhead

\noalign{\smallskip}\hline
\endfoot

\hline
\endlastfoot

\multicolumn{7}{c}{Metal-rich dwarfs (MRD)}\\
\hline\noalign{\smallskip}

Sun            &     0.00 $\pm$ 0.00 & 249 & 5768 $\pm$ 58  & 4.45 $\pm$ 0.12 & ...  $\pm$ ...  & 1.31 $\pm$ 0.09 & $-$0.01 $\pm$ 0.04 \\
HD~3765        & $-$63.18 $\pm$ 0.50 & 138 & 5310 $\pm$ 81  & 4.63 $\pm$ 0.24 & 4.55 $\pm$ 0.02 & 1.43 $\pm$ 0.10 & $+$0.12 $\pm$ 0.03 \\
HD~100167      & $-$29.41 $\pm$ 0.52 &  98 & 5898 $\pm$ 57  & 4.47 $\pm$ 0.12 & 4.43 $\pm$ 0.02 & 1.27 $\pm$ 0.09 & $+$0.01 $\pm$ 0.04 \\
61~UMa         & $-$5.47  $\pm$ 0.46 & 248 & 5555 $\pm$ 59  & 4.59 $\pm$ 0.13 & 4.52 $\pm$ 0.02 & 1.36 $\pm$ 0.10 & $-$0.07 $\pm$ 0.03 \\
$\beta$~Vir    & $+$4.58  $\pm$ 0.65 & 345 & 6199 $\pm$ 63  & 4.23 $\pm$ 0.13 & 4.09 $\pm$ 0.02 & 1.53 $\pm$ 0.08 & $+$0.17 $\pm$ 0.04 \\
$\beta$~CVn    & $+$6.34  $\pm$ 0.49 & 241 & 5902 $\pm$ 61  & 4.41 $\pm$ 0.13 & 4.37 $\pm$ 0.03 & 1.22 $\pm$ 0.10 & $-$0.20 $\pm$ 0.04 \\
$\beta$~Com    & $+$5.41  $\pm$ 0.57 & 198 & 6000 $\pm$ 58  & 4.44 $\pm$ 0.12 & 4.42 $\pm$ 0.01 & 1.30 $\pm$ 0.08 & $+$0.04 $\pm$ 0.04 \\
\end{longtable}
\clearpage
\twocolumn

\clearpage
\onecolumn
\begin{longtable}{lccccccc}

\caption{\label{tab:par_stars_stepar_vis} Stellar atmospheric parameters of the selected sample under {\sc StePar} restricted to the optical
 and $\log{g}$ values obtained with PARAM assuming parallaxes from {\emph Gaia} DR2 and the Hipparcos mission.}\\
\hline\noalign{\smallskip}
Name & $\varv_{\rm r}$    & S/N & $T_{\rm eff}$ & $\log{g}$ & $\log{g}_{\rm PARAM}$ & $\xi$          & [Fe/H]\\
     & [km\,s$^{-1}$]     &     & [K]           & [dex]     & [dex]                 & [km\,s$^{-1}$] & [dex]  \\
\noalign{\smallskip}\hline\noalign{\smallskip}
\endfirsthead

\caption{Stellar atmospheric parameters of the selected sample under {\sc StePar} restricted to the optical and $\log{g}$ values obtained
with PARAM assuming parallaxes from {\emph Gaia} DR2 and the Hipparcos mission (cont.).}\\
\hline\noalign{\smallskip}
Name & $\varv_{\rm r}$    & S/N & $T_{\rm eff}$ & $\log{g}$ & $\log{g}_{\rm PARAM}$ & $\xi$          & [Fe/H]\\
     & [km\,s$^{-1}$]     &     & [K]           & [dex]     & [dex]                 & [km\,s$^{-1}$] & [dex]  \\
\noalign{\smallskip}\hline\noalign{\smallskip}
\endhead

\noalign{\smallskip}\hline
\endfoot

\hline
\endlastfoot

\multicolumn{7}{c}{Metal-rich dwarfs (MRD)}\\
\hline\noalign{\smallskip}

Sun            &     0.00 $\pm$ 0.00 & 249 & 5787 $\pm$ 54  & 4.42 $\pm$ 0.11 & ...  $\pm$ ...  & 0.98 $\pm$ 0.08 & $+$0.02 $\pm$ 0.03 \\
HD~3765        & $-$63.18 $\pm$ 0.50 & 138 & 5206 $\pm$ 84  & 4.62 $\pm$ 0.22 & 4.55 $\pm$ 0.02 & 1.17 $\pm$ 0.12 & $+$0.16 $\pm$ 0.04 \\
HD~100167      & $-$29.41 $\pm$ 0.52 &  98 & 5942 $\pm$ 54  & 4.49 $\pm$ 0.10 & 4.43 $\pm$ 0.02 & 1.04 $\pm$ 0.09 & $+$0.04 $\pm$ 0.04 \\
61~UMa         & $-$5.47  $\pm$ 0.46 & 248 & 5576 $\pm$ 56  & 4.56 $\pm$ 0.12 & 4.52 $\pm$ 0.02 & 1.07 $\pm$ 0.09 & $-$0.01 $\pm$ 0.03 \\
$\beta$~Vir    & $+$4.58  $\pm$ 0.65 & 345 & 6255 $\pm$ 64  & 4.20 $\pm$ 0.12 & 4.09 $\pm$ 0.02 & 1.45 $\pm$ 0.08 & $+$0.23 $\pm$ 0.04 \\
$\beta$~CVn    & $+$6.34  $\pm$ 0.49 & 241 & 5967 $\pm$ 57  & 4.40 $\pm$ 0.12 & 4.37 $\pm$ 0.03 & 0.96 $\pm$ 0.09 & $-$0.14 $\pm$ 0.04 \\
$\beta$~Com    & $+$5.41  $\pm$ 0.57 & 198 & 6105 $\pm$ 56  & 4.42 $\pm$ 0.11 & 4.42 $\pm$ 0.01 & 1.18 $\pm$ 0.07 & $+$0.02 $\pm$ 0.04 \\
\end{longtable}
\clearpage
\twocolumn

\clearpage
\onecolumn
\begin{longtable}{cccccccc}

\caption{\label{tab:line_table_all_fe_i} Merged \ion{Fe}{i} linelists.}\\
\hline\noalign{\smallskip}
${\lambda}_{\rm air}$ & ${\chi}_{\rm I}$ & $\log{gf}$ & \multicolumn{4}{c}{Linelist$^a$} & Reference$^b$\\
{[}\AA{]} & [eV] & & MRD & MPD & MRG & MPG & \\
\noalign{\smallskip}\hline\noalign{\smallskip}
\endfirsthead

\caption{Merged \ion{Fe}{i} linelists (cont.).}\\
\hline\noalign{\smallskip}
${\lambda}_{\rm air}$ & ${\chi}_{\rm I}$ & $\log{gf}$ &\multicolumn{4}{c}{Linelist$^a$} & Reference$^b$\\
{[}\AA{]} & [eV] & & MRD & MPD & MRG & MPG & \\
\noalign{\smallskip}\hline\noalign{\smallskip}
\endhead

\hline
\endfoot

\hline
\noalign{\smallskip}
\multicolumn{8}{p{10cm}}{$^a$\,\footnotesize {\bf Linelists.} MRD: metal-rich dwarfs; MPD: metal-poor dwarfs; MRG: metal-rich giants; MPG: metal-poor giants.}\\
\multicolumn{8}{p{10cm}}{$^b$\,\footnotesize {\bf References.} Sou08: \citet{Sou08}; And16: \citet{And16}.}\\
\endlastfoot

 5307.361 & 1.61 & $-$2.912 & \textbullet & \textbullet &             & \textbullet & Sou08 \\
 5321.108 & 4.44 & $-$1.089 &             &             &             & \textbullet & Sou08 \\
 5322.041 & 2.28 & $-$2.802 & \textbullet &             &             & \textbullet & Sou08 \\
 5339.929 & 3.27 & $-$0.635 &             &             &             & \textbullet & Sou08 \\
 5364.871 & 4.45 & $+$0.228 &             &             & \textbullet & \textbullet & Sou08 \\
 5373.709 & 4.47 & $-$0.710 & \textbullet & \textbullet &             &             & Sou08 \\
 5379.574 & 3.70 & $-$1.514 & \textbullet & \textbullet & \textbullet & \textbullet & Sou08 \\
 5386.333 & 4.15 & $-$1.670 & \textbullet &             & \textbullet & \textbullet & Sou08 \\
\end{longtable}
\clearpage
\twocolumn

\clearpage
\onecolumn
\begin{longtable}{cccccccc}

\caption{\label{tab:line_table_all_fe_ii} Merged \ion{Fe}{ii} linelists.}\\
\hline\noalign{\smallskip}
${\lambda}_{\rm air}$ & ${\chi}_{\rm II}$ & $\log{gf}$ & \multicolumn{4}{c}{Linelist$^a$} & Reference$^b$\\
{[}{\AA}{]} & [eV] & & MRD & MPD & MRG & MPG & \\
\noalign{\smallskip}\hline\noalign{\smallskip}
\endfirsthead

\caption{Merged \ion{Fe}{ii} linelists (cont.).}\\
\hline\noalign{\smallskip}
${\lambda}_{\rm air}$ & ${\chi}_{\rm II}$ & $\log{gf}$ & \multicolumn{4}{c}{Linelist$^a$} & Reference$^b$\\
{[}{\AA}{]} & [eV] & & MRD & MPD & MRG & MPG & \\
\noalign{\smallskip}\hline\noalign{\smallskip}
\endhead

\hline
\endfoot

\hline
\noalign{\smallskip}
\multicolumn{8}{p{10cm}}{$^a$\,\footnotesize {\bf Linelists.} MRD: metal-rich dwarfs; MPD: metal-poor dwarfs; MRG: metal-rich giants; MPG: metal-poor giants.}\\
\multicolumn{8}{p{10cm}}{$^b$\,\footnotesize {\bf References.} Sou08: \citet{Sou08}; And16: \citet{And16}.}\\
\endlastfoot

 5325.552 & 3.22 & $-$3.160 &             & \textbullet &             & \textbullet & Sou08 \\
 5414.070 & 3.22 & $-$3.580 &             &             & \textbullet &             & Sou08 \\
 5425.248 & 3.20 & $-$3.220 & \textbullet & \textbullet & \textbullet & \textbullet & Sou08 \\
 5534.838 & 3.24 & $-$2.730 &             & \textbullet &             &             & Sou08 \\
 5991.371 & 3.15 & $-$3.540 & \textbullet &             &             &             & Sou08 \\
 6084.102 & 3.20 & $-$3.780 & \textbullet &             & \textbullet &             & Sou08 \\
 6149.246 & 3.89 & $-$2.720 & \textbullet &             & \textbullet & \textbullet & Sou08 \\
 6238.386 & 3.89 & $-$2.754 &             & \textbullet & \textbullet &             & Sou08 \\
\end{longtable}
\clearpage
\twocolumn

\begin{landscape}
\begin{figure}
\centering
\includegraphics{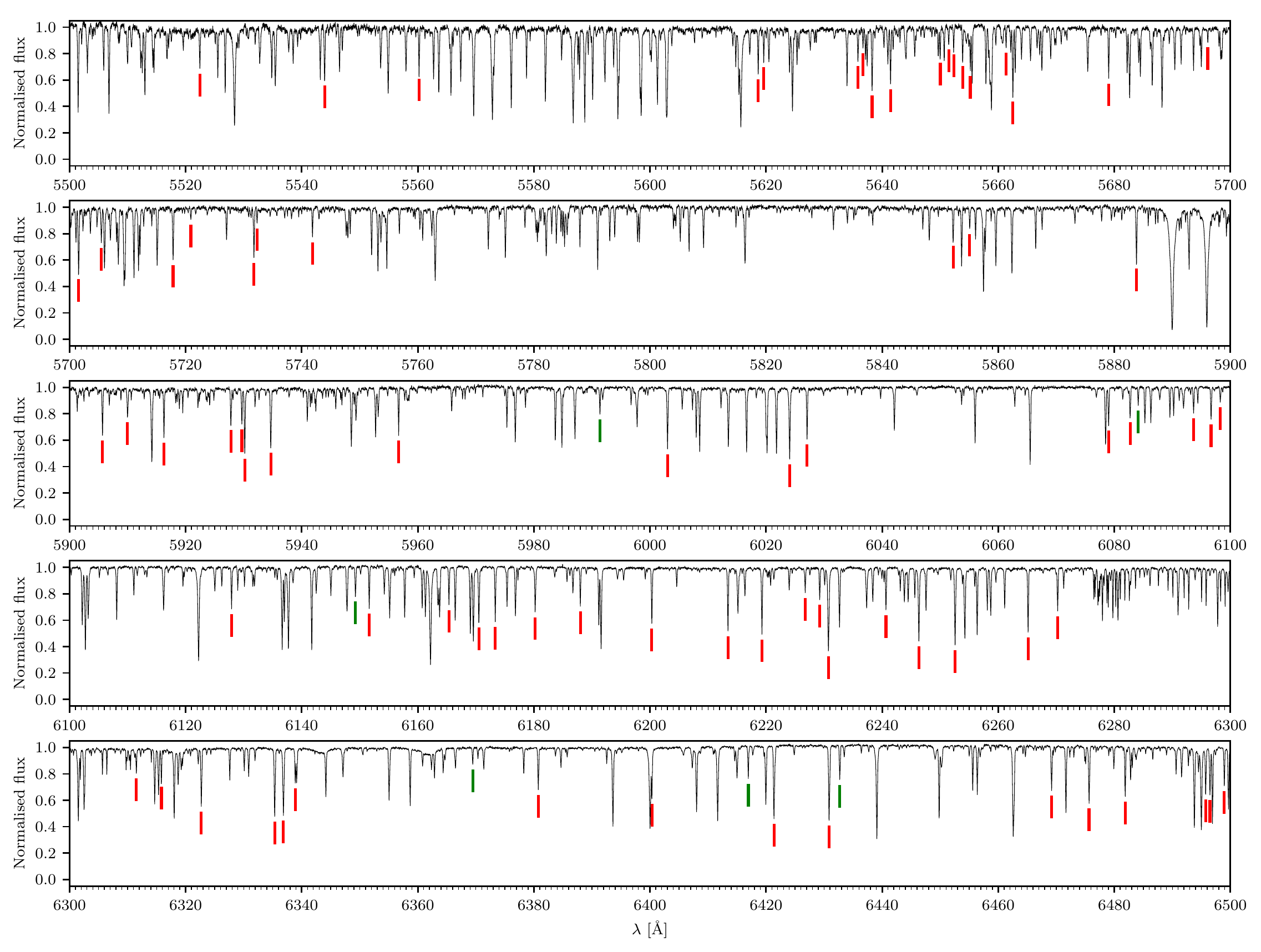}
\caption{\label{fig:18Scospectrum} CARMENES spectrum of 18 Sco. \ion{Fe}{i} and \ion{Fe}{ii} lines are shown in red and green, respectively (cont.).}
\end{figure}
\end{landscape}

\begin{landscape}
\begin{figure}
\centering
\includegraphics{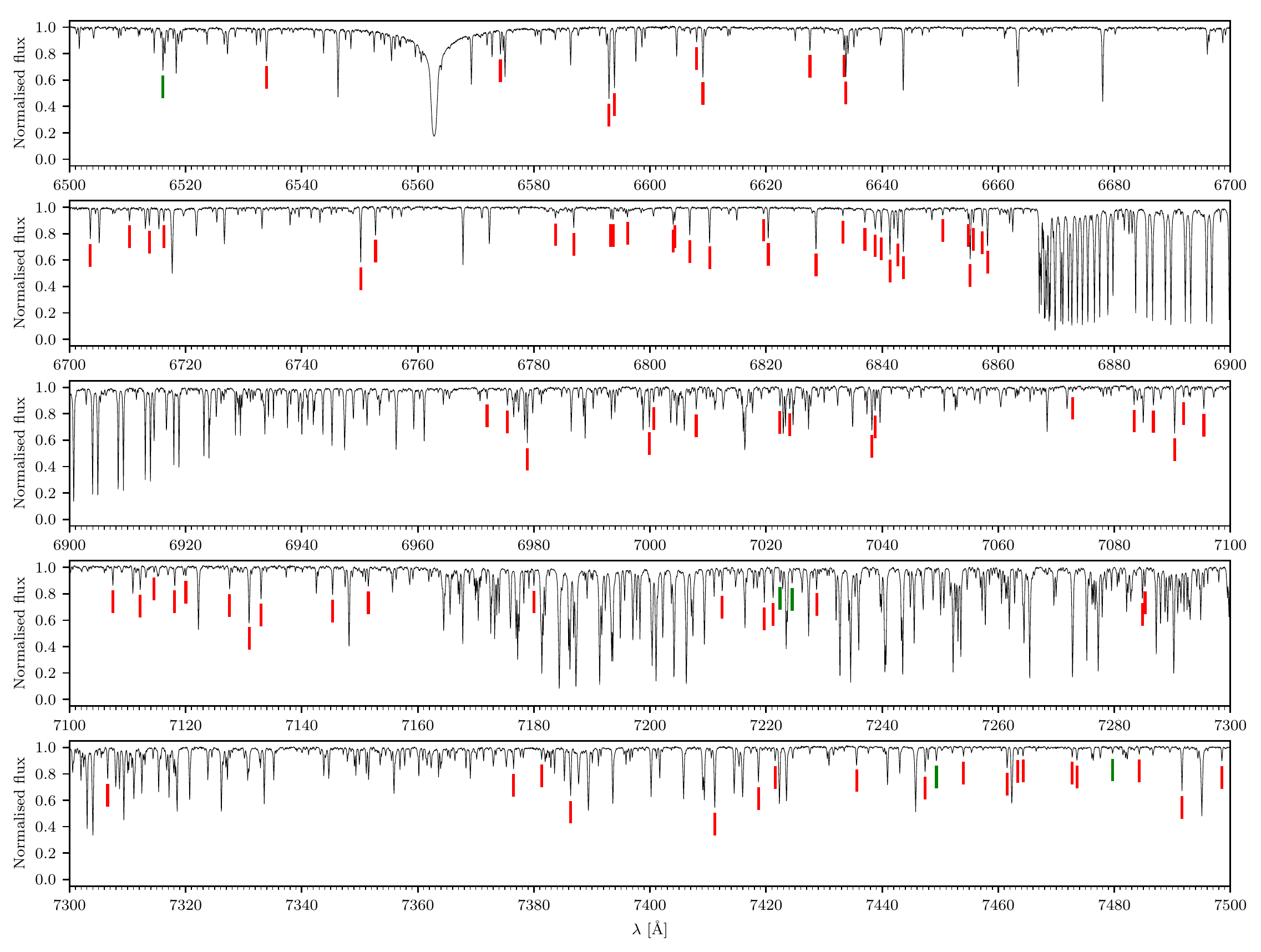}
\contcaption{CARMENES spectrum of 18 Sco.}
\end{figure}
\end{landscape}

\begin{landscape}
\begin{figure}
\centering
\includegraphics{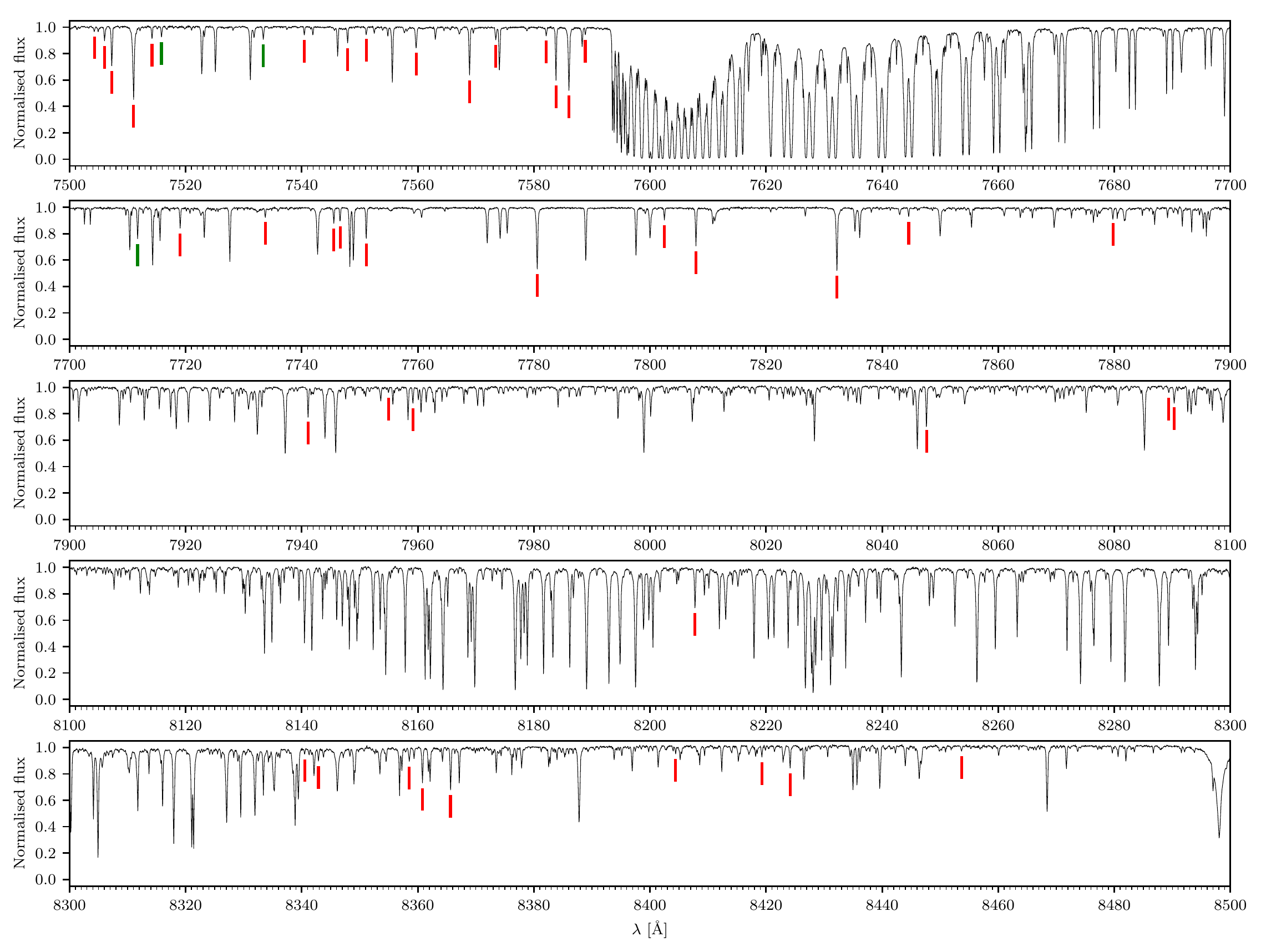}
\contcaption{CARMENES spectrum of 18 Sco.}
\end{figure}
\end{landscape}

\begin{landscape}
\begin{figure}
\centering
\includegraphics{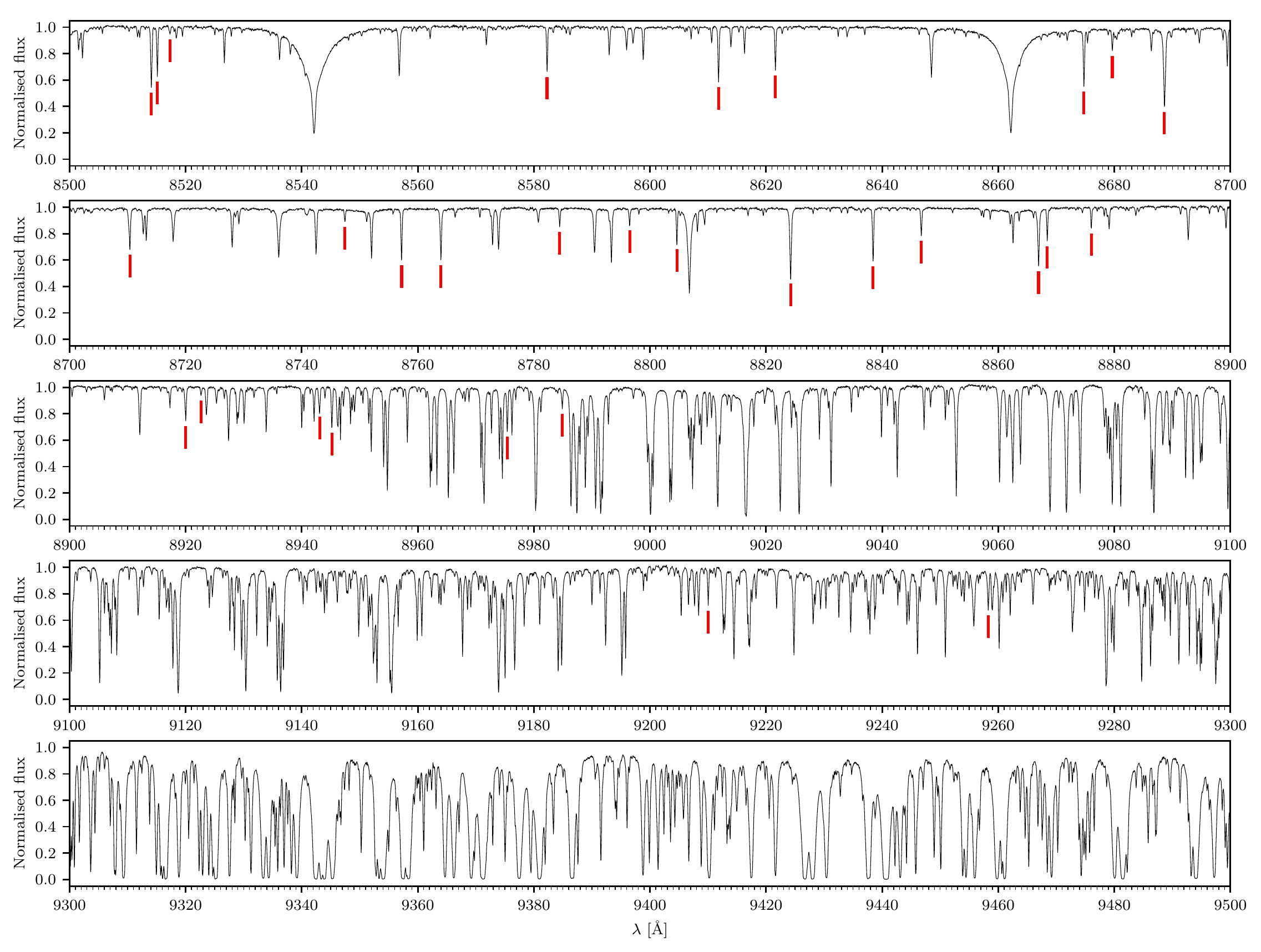}
\contcaption{CARMENES spectrum of 18 Sco.}
\end{figure}
\end{landscape}

\begin{landscape}
\begin{figure}
\centering
\includegraphics{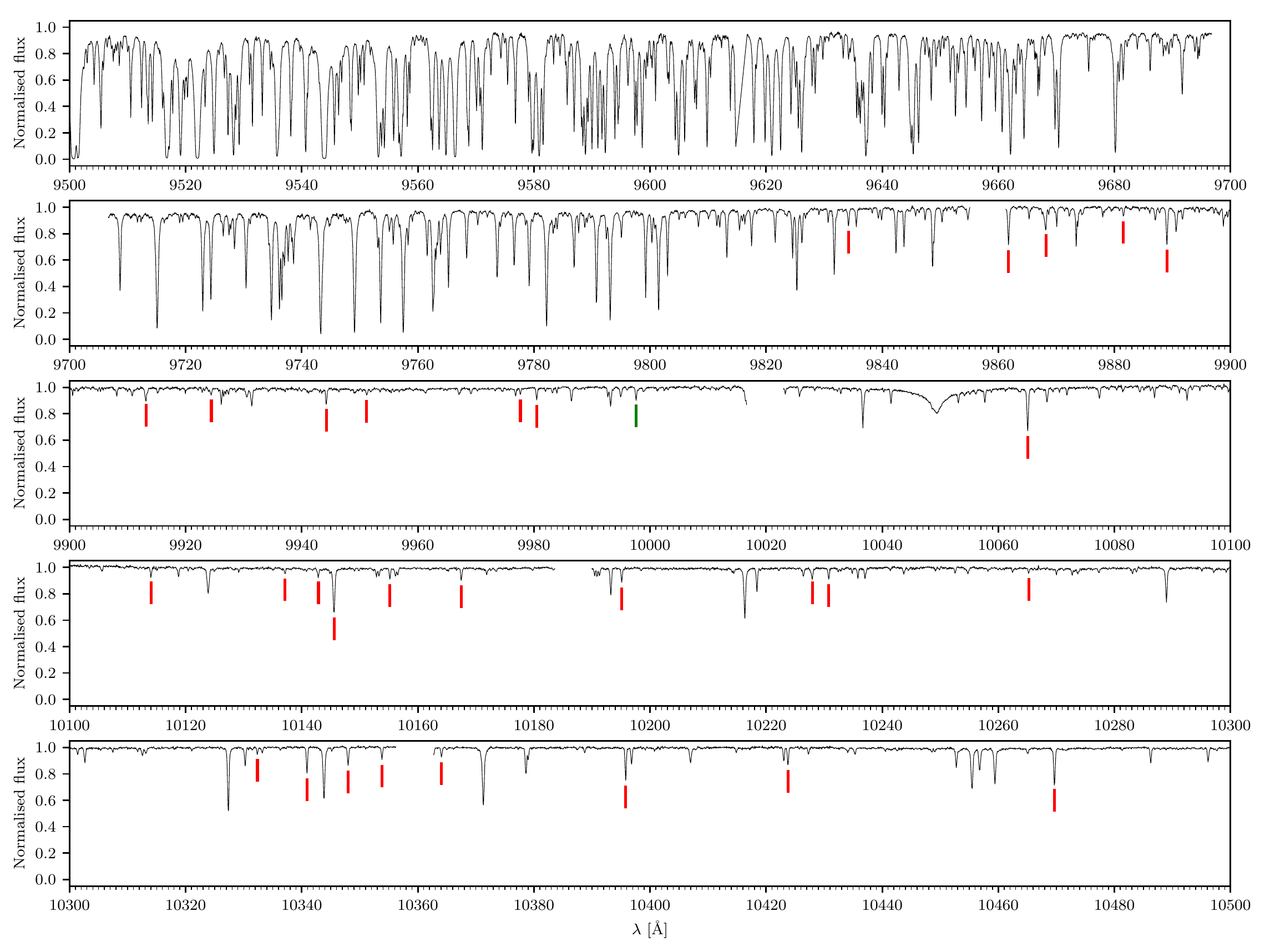}
\contcaption{CARMENES spectrum of 18 Sco.}
\end{figure}
\end{landscape}

\begin{landscape}
\begin{figure}
\centering
\includegraphics{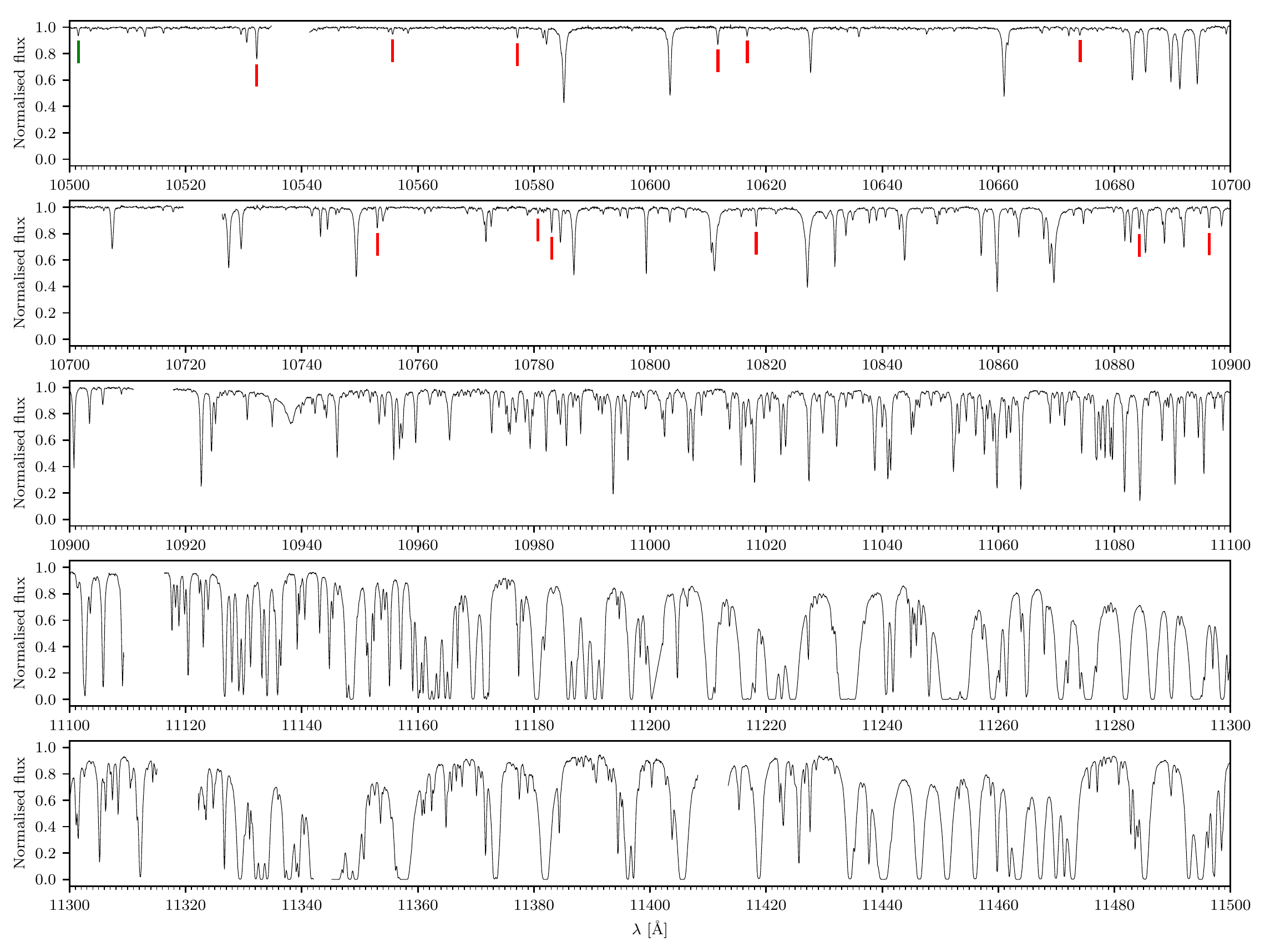}
\contcaption{CARMENES spectrum of 18 Sco.}
\end{figure}
\end{landscape}

\begin{landscape}
\begin{figure}
\centering
\includegraphics{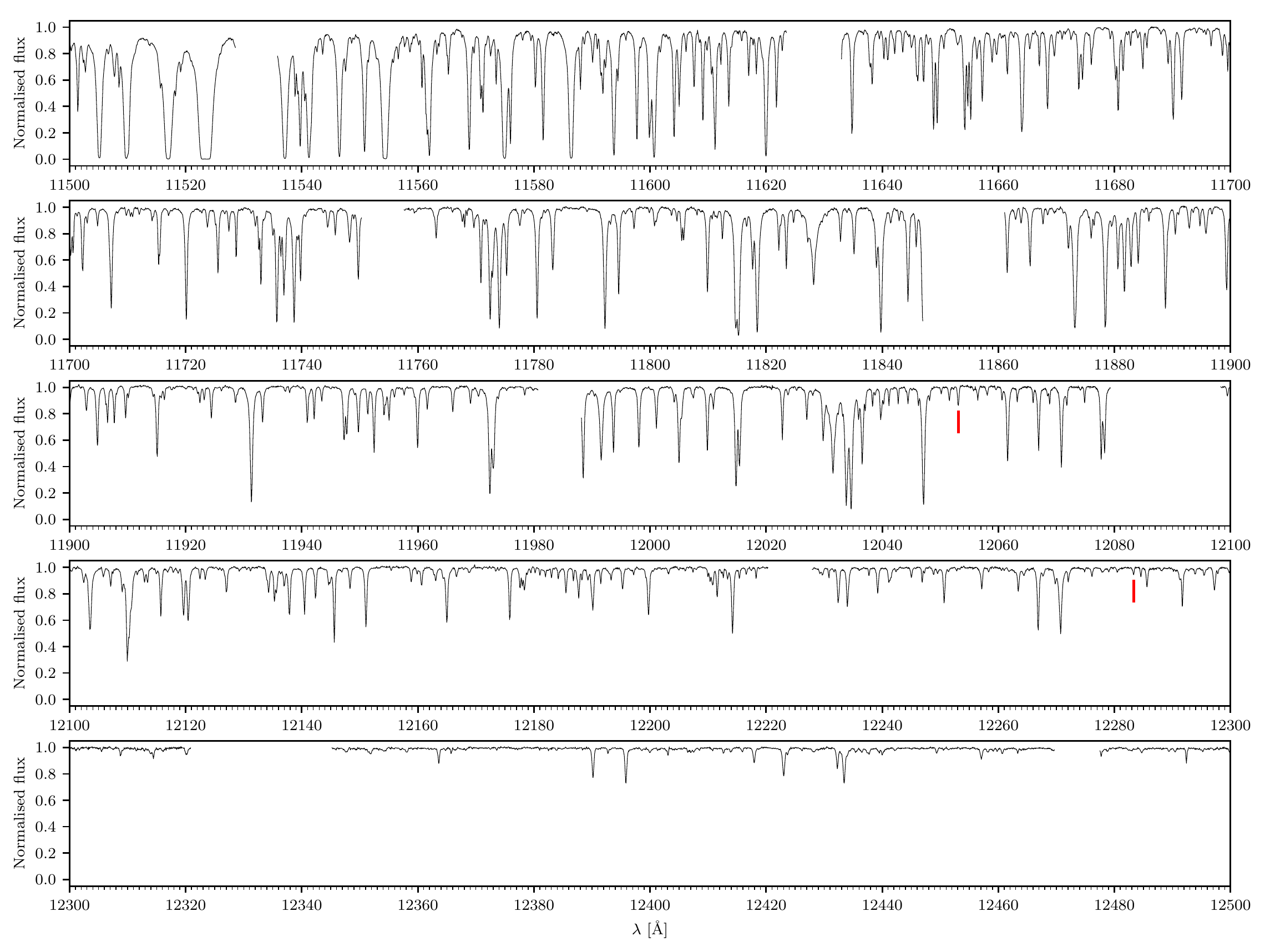}
\contcaption{CARMENES spectrum of 18 Sco.}
\end{figure}
\end{landscape}

\begin{landscape}
\begin{figure}
\centering
\includegraphics{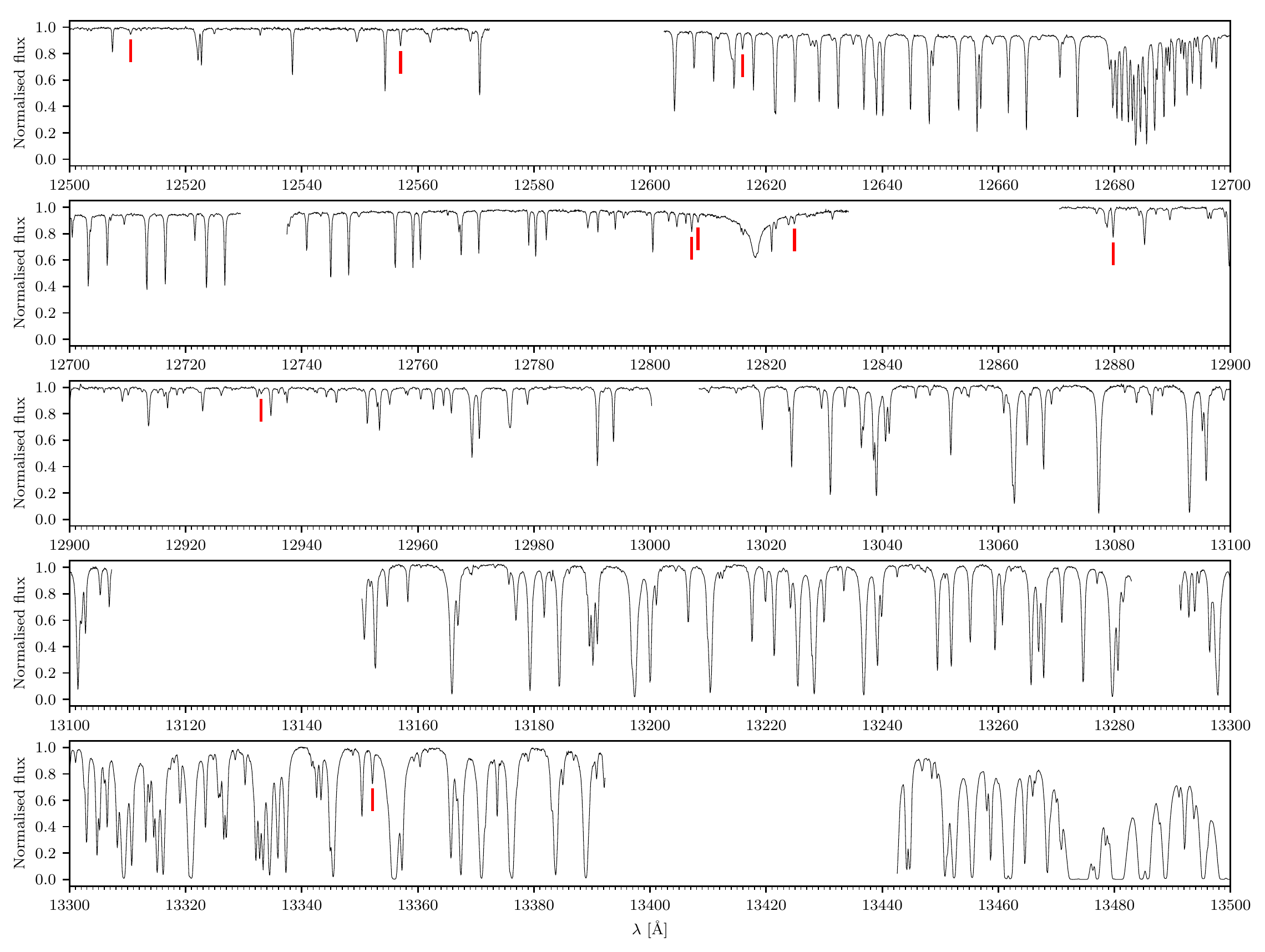}
\contcaption{CARMENES spectrum of 18 Sco.}
\end{figure}
\end{landscape}

\begin{landscape}
\begin{figure}
\centering
\includegraphics{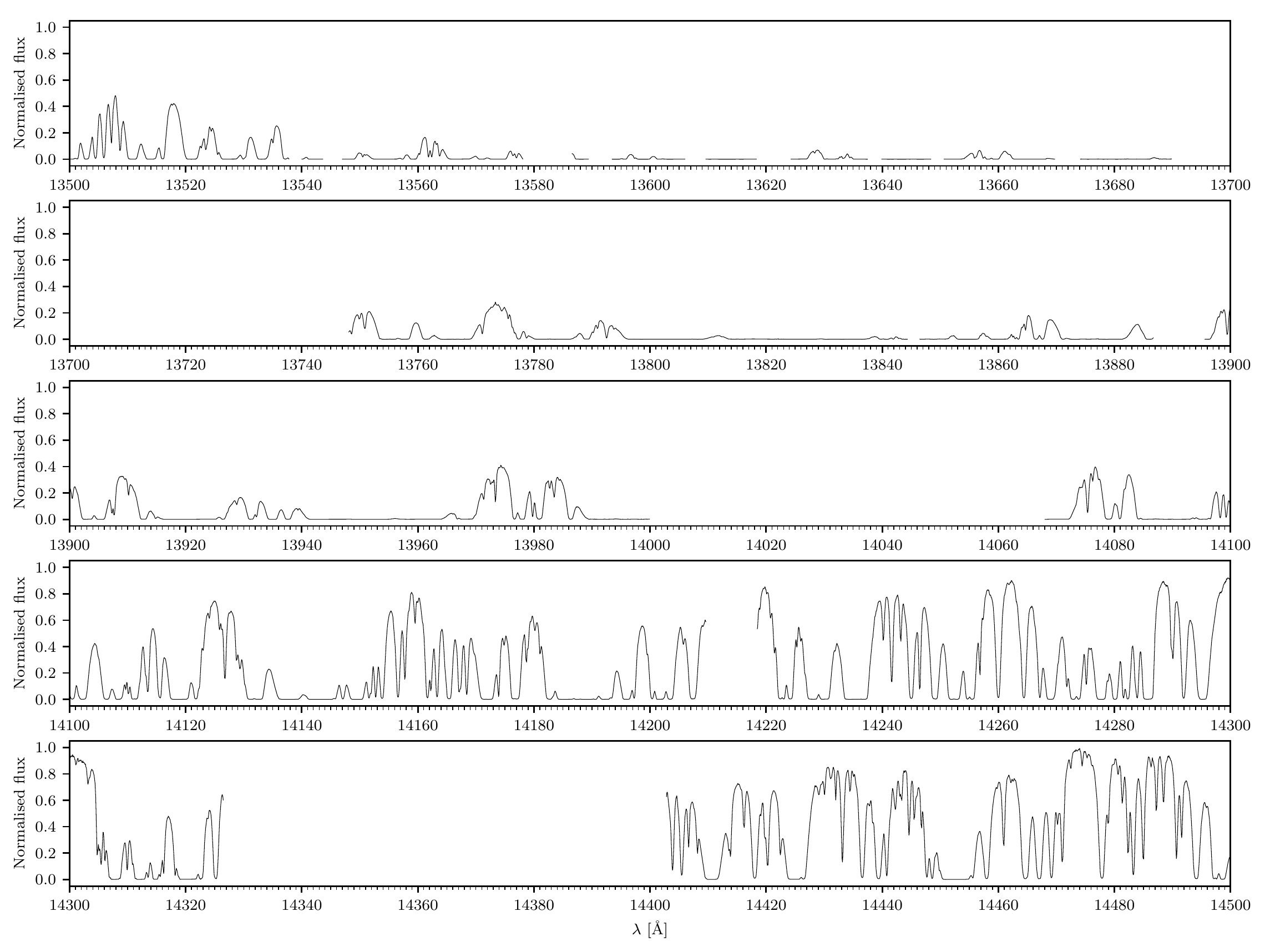}
\contcaption{CARMENES spectrum of 18 Sco.}
\end{figure}
\end{landscape}

\begin{landscape}
\begin{figure}
\centering
\includegraphics{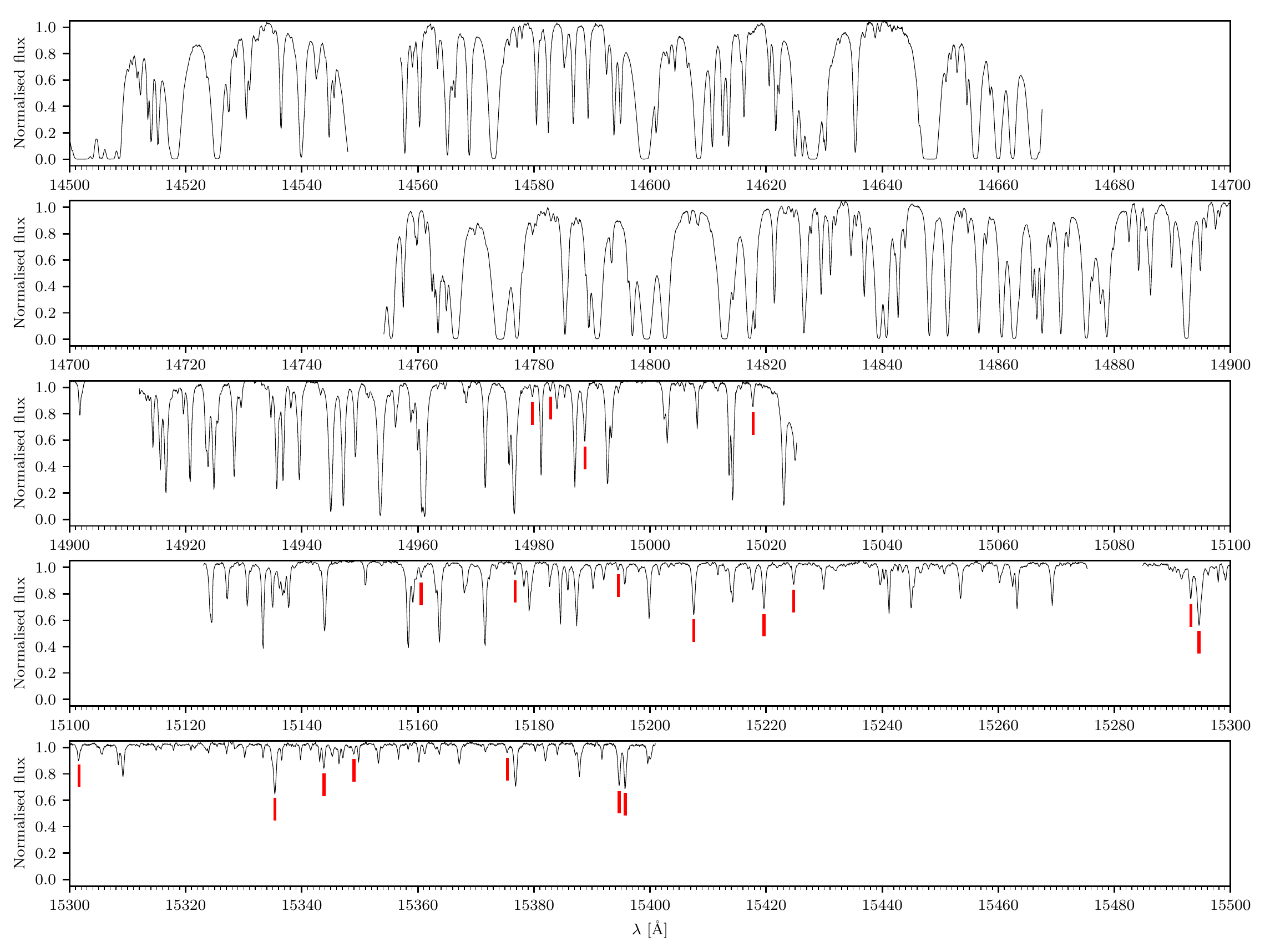}
\contcaption{CARMENES spectrum of 18 Sco.}
\end{figure}
\end{landscape}

\begin{landscape}
\begin{figure}
\centering
\includegraphics{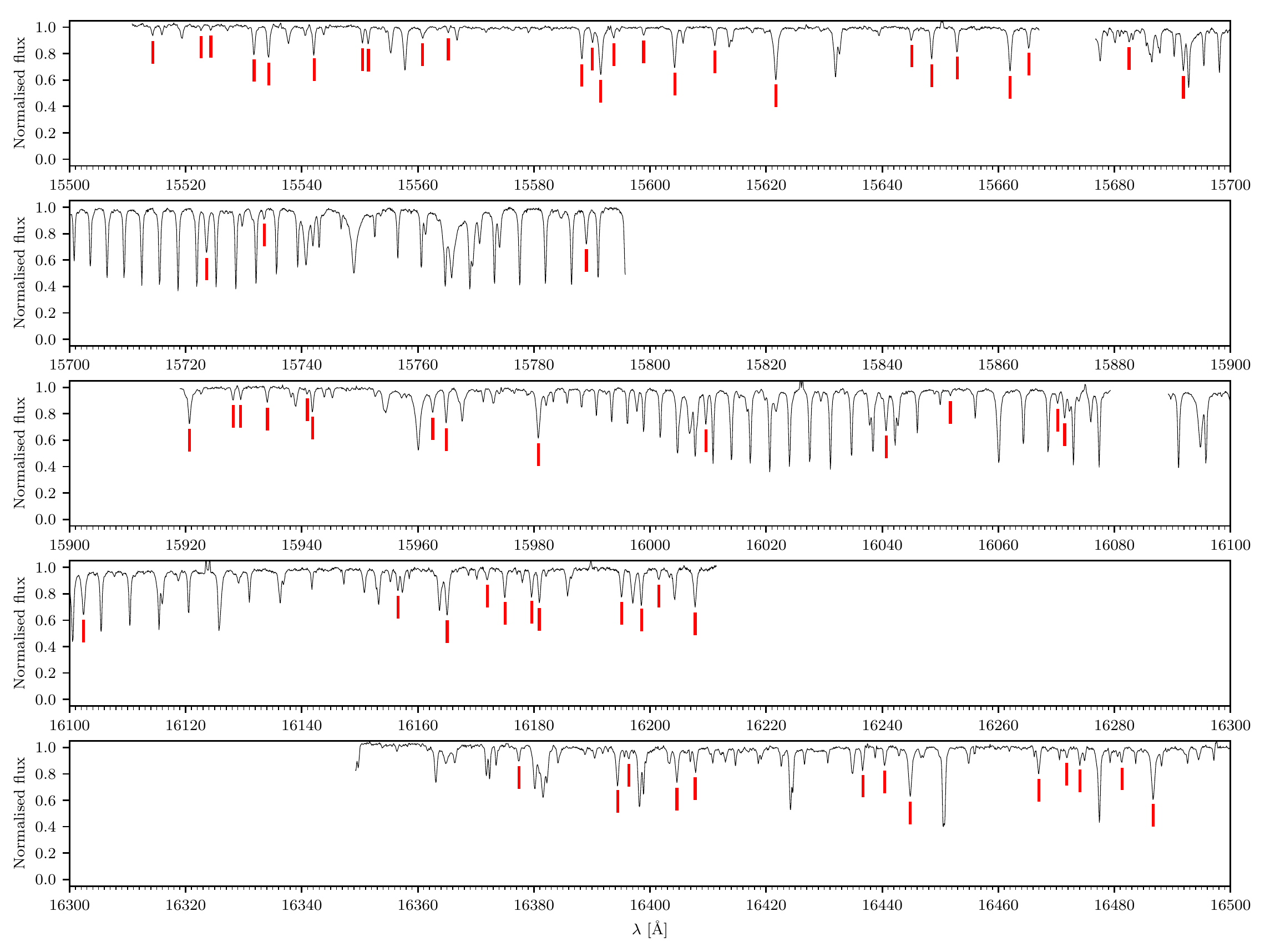}
\contcaption{CARMENES spectrum of 18 Sco.}
\end{figure}
\end{landscape}

\begin{landscape}
\begin{figure}
\centering
\includegraphics{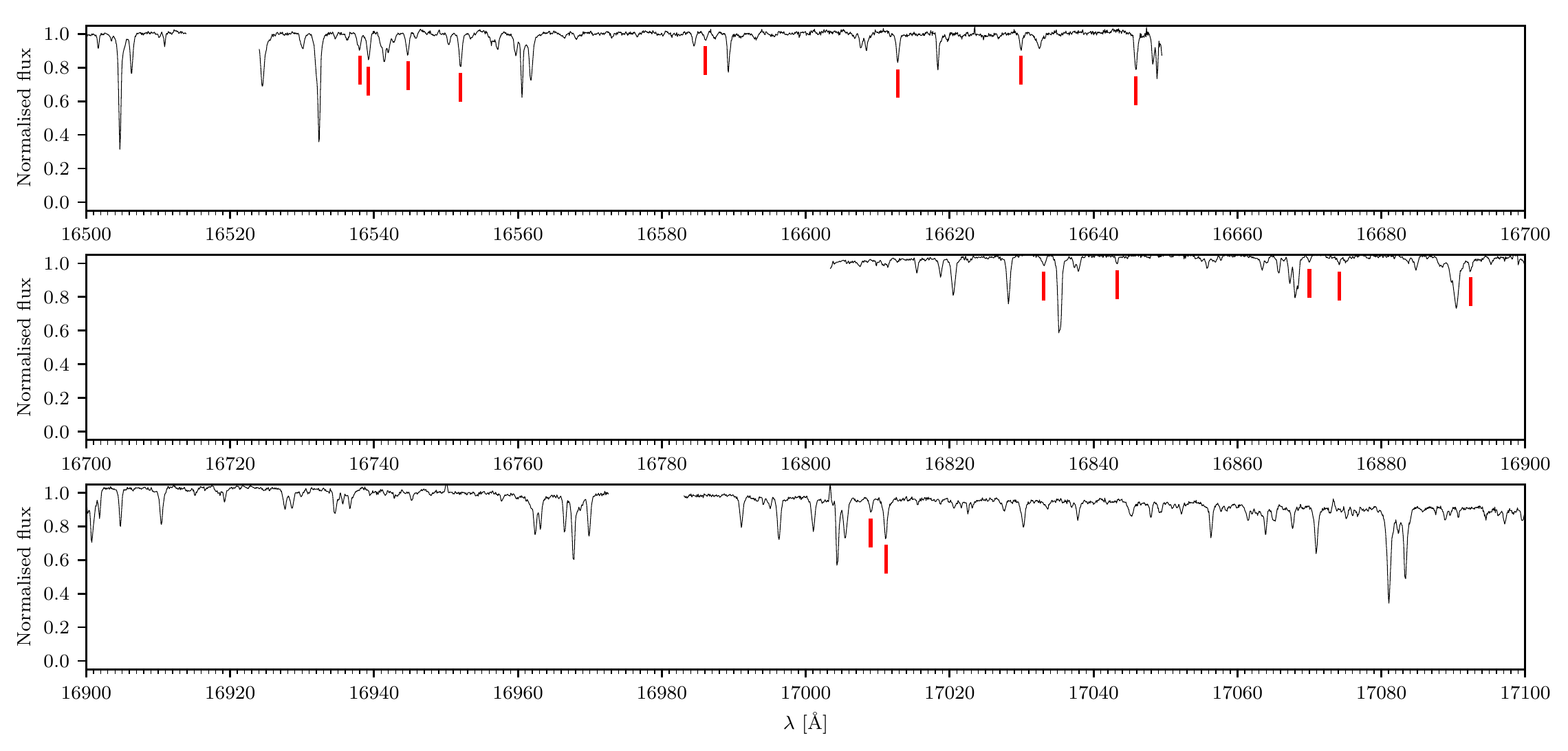}
\contcaption{CARMENES spectrum of 18 Sco.}
\end{figure}
\end{landscape}


\bsp 
\label{lastpage}
\end{document}